\documentclass[twocolumn]{aastex62}

\hypersetup{linkcolor=blue,citecolor=blue,filecolor=cyan,urlcolor=blue}

\usepackage{braket}
\usepackage{amsmath}
\usepackage{bm}

\newcommand{\lya}{Ly$\alpha\,$}

\newcommand{\cii}{[\ion{C}{2}]}

\newcommand{\mgii}{\ion{Mg}{2}}

\defcitealias{E20prep}{Paper II}
\submitjournal{ApJ}

\shorttitle{Detecting and Characterizing Young Quasars I}
\shortauthors{A.-C. Eilers et al.}

\begin{document}\frenchspacing

\title{\textbf{Detecting and Characterizing Young Quasars I: Systemic Redshifts and\\ Proximity Zones Measurements}}

\author[0000-0003-2895-6218]{Anna-Christina Eilers}\thanks{NASA Hubble Fellow}
\affiliation{MIT Kavli Institute for Astrophysics and Space Research, 77 Massachusetts Ave., Cambridge, MA 02139, USA}
\affiliation{Max-Planck-Institute for Astronomy, K\"onigstuhl 17, 69117 Heidelberg, Germany}

\author[0000-0002-7054-4332]{Joseph F. Hennawi}
\affiliation{Physics Department, University of California, Santa Barbara, CA 93106-9530, USA}
\affiliation{Max-Planck-Institute for Astronomy, K\"onigstuhl 17, 69117 Heidelberg, Germany}

\author[0000-0002-2662-8803]{Roberto Decarli}
\affiliation{INAF -- Osservatorio di Astrofisica e Scienza dello Spazio di Bologna, via Gobetti 93/3, I-40129, Bologna, Italy}

\author[0000-0003-0821-3644]{Frederick B. Davies}
\affiliation{Physics Department, University of California, Santa Barbara, CA 93106-9530, USA}
\affiliation{Lawrence Berkeley National Laboratory, CA 94720-8139, USA}

\author[0000-0001-9024-8322]{Bram Venemans}
\affiliation{Max-Planck-Institute for Astronomy, K\"onigstuhl 17, 69117 Heidelberg, Germany}

\author[0000-0003-4793-7880]{Fabian Walter}
\affiliation{Max-Planck-Institute for Astronomy, K\"onigstuhl 17, 69117 Heidelberg, Germany}

\author[0000-0002-2931-7824]{Eduardo Ba{\~n}ados}
\affiliation{Max-Planck-Institute for Astronomy, K\"onigstuhl 17, 69117 Heidelberg, Germany}

\author[0000-0003-3310-0131]{Xiaohui Fan}
\affiliation{Steward Observatory, University of Arizona, 933 North Cherry Avenue, Tucson, AZ 85721, USA}

\author[0000-0002-6822-2254]{Emanuele P. Farina}
\affiliation{Max Planck Institut f\"ur Astrophysik, Karl--Schwarzschild--Stra{\ss}e 1, D-85748 Garching bei M\"unchen, Germany}

\author[0000-0002-5941-5214]{Chiara Mazzucchelli}
\affiliation{European Southern Observatory, Alonso de Cordova 3107, Vitacura, Region Metropolitana, Chile}

\author[0000-0001-8695-825X]{Mladen Novak}
\affiliation{Max-Planck-Institute for Astronomy, K\"onigstuhl 17, 69117 Heidelberg, Germany}

\author[0000-0002-4544-8242]{Jan-Torge Schindler}
\affiliation{Max-Planck-Institute for Astronomy, K\"onigstuhl 17, 69117 Heidelberg, Germany}

\author[0000-0003-3769-9559]{Robert A. Simcoe}
\affiliation{MIT Kavli Institute for Astrophysics and Space Research, 77 Massachusetts Ave., Cambridge, MA 02139, USA}

\author[0000-0002-7633-431X]{Feige Wang}\thanks{NASA Hubble Fellow}
\affiliation{Steward Observatory, University of Arizona, 933 North Cherry Avenue, Tucson, AZ 85721, USA}

\author[0000-0001-5287-4242]{Jinyi Yang}
\affiliation{Steward Observatory, University of Arizona, 933 North Cherry Avenue, Tucson, AZ 85721, USA}

\correspondingauthor{Anna-Christina Eilers}
\email{eilers@mit.edu}

\begin{abstract}
In a multi-wavelength survey of $13$ quasars at $5.8\lesssim z\lesssim6.5$, that were pre--selected to be potentially young, we find five objects with extremely small proximity zone sizes that may imply UV--luminous quasar lifetimes of $\lesssim 100,000$ years. Proximity zones are regions of enhanced transmitted flux in the vicinity of the quasars that are sensitive to the quasars' lifetimes because the intergalactic gas has a finite response time to their radiation. We combine sub-mm observations from the Atacama Large Millimetre Array (ALMA) and the NOrthern Extended Millimeter Array (NOEMA), as well as deep optical and near-infrared spectra from medium-resolution spectrograph on the Very Large Telescope (VLT) and on the Keck telescopes, in order to identify and characterize these new young quasars, which provide valuable clues about the accretion behavior of supermassive black holes (SMBHs) in the early universe, and pose challenges on current black hole formation models to explain the rapid formation of billion solar mass black holes. We measure the quasars' systemic redshifts, black hole masses, Eddington ratios, emission line luminosities, and star formation rates of their host galaxies. Combined with previous results we estimate the fraction of young objects within the high--redshift quasar population at large to be $5\%\lesssim f_{\rm young}\lesssim 10\%$. One of the young objects, PSO\,J158--14, shows a very bright dust continuum flux ($F_{\rm cont}=3.46\pm 0.02\,\rm mJy$), indicating a highly star--bursting host galaxy with a star formation rate of approximately $1420\,M_{\odot}\,\rm yr^{-1}$. 
\end{abstract}

\keywords{dark ages, early universe --- quasars: emission lines, supermassive black holes --- methods: data analysis --- intergalactic medium --- submillimeter: ISM, galaxies} 

\section{Introduction}

High-redshift quasars host central supermassive black holes (SMBHs) with masses exceeding $M_{\rm BH}\sim 10^9-10^{10}~M_{\odot}$ as early as $\lesssim 1$~Gyr after the Big Bang \citep[e.g.][]{Mortlock2011, Venemans2013, Wu2015, Mazzucchelli2017, Banados2018, Onoue2019}. How these SMBHs form and grow in such short amounts of cosmic time remains an unanswered question.
Assuming Eddington limited accretion rates and a constant supply of fueling material SMBHs grow exponentially during the quasar's lifetime $t_{\rm Q}$, i.e. 
\begin{equation}
    M_{\rm BH}(t_{\rm Q})=M_{\rm seed}\cdot \exp\left(\frac{t_{\rm Q}}{t_{\rm S}}\right).  \label{eq:mbh}
\end{equation}
The initial mass $M_{\rm seed}$ denotes the mass of the black hole before the onset of quasar activity. The lifetime or the age of a quasar $t_{\rm Q}$ is defined such that the onset of the quasar activity happened at a time $-t_{\rm Q}$ in the past. 
The e-folding time, or ``Salpeter'' time, $t_{\rm S}$ \citep{Salpeter1964}, describes the characteristic time scale on which the black hole growth is believed to occur, i.e. 
\begin{equation}
    t_{\rm S} \simeq 4.5\times10^7 \left(\frac{\epsilon}{0.1}\right)\left(\frac{L_{\rm bol}}{L_{\rm edd}}\right)^{-1}~\rm yr, 
\end{equation}
where $\epsilon$ denotes the radiative efficiency of the accretion, which is assumed to be about $10\%$ in thin accretion disk models \citep{ShakuraSunyaev1973}, and $L_{\rm bol}$ describes the bolometric luminosity of the quasar with a theoretical upper limit of the Eddington luminosity $L_{\rm edd}$. It requires at least $16$ e-foldings, i.e. $\gtrsim 7 \times10^8$~yr, in order to grow a billion solar mass black hole from an initial stellar remnant black hole seed with $M_{\rm seed}\sim 100\,M_{\odot}$, even if they accrete continuously at the Eddington limit \citep[e.g.][]{Volonteri2010, Volonteri2012}. 
However, it is currently unknown whether quasars obey this exponential light curve, or if other physics related to the triggering of quasar activity and the supply of fuel complicate this simple picture, giving rise to much more complex light curves \citep[e.g.][]{DiMatteo2005, Springel2005, Hopkins2005, Novak2011, Davies2019a}.

These timescales required for the growth of SMBHs are comparable to the age of the universe at $z\gtrsim 6$. Nevertheless, at these high redshifts more than $200$ quasars have been discovered in the last decade \citep[e.g.][]{Venemans2015, Banados2016, Mazzucchelli2017, Wang2019, Yang2019, Reed2019}, many of which host billion solar mass black holes. Thus, massive initial seeds in excess of stellar remnants, i.e. $M_{\rm seed} \gtrsim 1000\,M_{\odot}$ \citep[e.g.][]{LodatoNatarajan2006, Visbal2014, Habouzit2016, Schauer2017}, or radiatively  inefficient accretion rates with $\epsilon \lesssim 0.01-0.001$ have been invoked \citep[e.g.][]{Volonteri2015, Trakthenbrot2017, Davies2019b}, which would reduce the quasar lifetime required to grow the SMBHs. 

Measurements of quasar lifetimes have proven to be challenging. At low redshifts, i.e. $z\sim 2-4$, the quasar lifetime can be constrained by comparing the number density of quasars to their host dark matter halo abundance inferred from clustering studies \citep{HaimanHui2001, MartiniWeinberg2001, Martini2004, WhiteMartiniCohn2008}. However, to date this method has yielded only weak constraints on $t_{\rm Q} \sim 10^6-10^9$~yr owing to uncertainties in how quasars populate dark matter halos \citep{Shen2009, White2012, ConroyWhite2013, Cen2015}. Following the ``Soltan'' argument \citep{Soltan1982}, which states that the luminosity function of quasars as a function of redshift reflects the gas accretion history of local remnant black holes, \citet{YuTremaine2002} estimated the mean lifetime of luminous quasars from local early-type galaxies to be $t_{\rm Q}\sim 10^7-10^8$~yr. Further constraints on quasar activity on timescales between $\sim 10^5-10^7$~yr have been set by measuring an ionization ``echo'' of the quasar, which denotes the time-lag between changes in the quasar's ionization rate and the responding changes in the opacity of the surrounding IGM \citep{Adelberger2004, Hennawi2006, Schmidt2017, Bosman2019}. 
%An upper limit on the quasar lifetime, $t_{\rm Q} < 10^9$~yr, is set by the observed evolution of the quasar luminosity function, since the whole quasar population rises and falls over roughly this timescale \citep{Osmer1998}. 
A recent compilation of studies on the timescales governing the growth of SMBHs can be found in \citet{araa_smbh2020}.

We recently showed how the extent of the proximity zones around high-redshift quasars provides a new and independent constraint on the lifetime of quasars \citep{Eilers2017a, Eilers2018b, Khrykin2019, Davies2019b}. These regions of enhanced transmitted flux within the \lya\ forest in the immediate vicinity of the quasars have been ionized by the quasar's intense radiation itself \citep[e.g.][]{Bajtlik1988, HaimanCen2001, Wyithe2005, BoltonHaehnelt2007b, Lidz2007, Bolton2011, Keating2015}, and are sensitive to the lifetime of the quasars because intergalactic gas has a finite response time to the quasars' radiation \citep[e.g.][]{Khrykin2016, Eilers2017a, Davies2019a}. The equilibration timescale $t_{\rm eq}$ describes the time when the intergalactic medium (IGM) has reached ionization equilibrium with the ionizing photons emitted by the quasar, i.e. $t_{\rm eq} \approx \Gamma^{-1}_{\rm HI}$, where $\Gamma_{\rm HI}$ denotes the total photoionization rate from the quasar as well as the ultraviolet background (UVB). However, the quasar's radiation dominates the radiation field within the proximity zone, which has been observationally defined as the location at which the smoothed, continuum-normalized transmitted flux drops below the $10\%$-level \citep{Fan2006}. A photoionization rate of $\Gamma_{\rm HI}\approx 10^{-12}\,\rm s^{-1}$ from the quasar's radiation at the ``edge'' of the proximity zone at $z\approx 6$ leads to an equilibration timescale of $t_{\rm eq}\approx 3\times 10^4$~yr \citep{Davies2019a}. 

Applying this new method to a data set of $31$ quasar spectra at $5.8\lesssim z\lesssim 6.5$ \citep{Eilers2018a}, we discovered an unexpected population of quasars with significantly smaller proximity zones than expected, that are likely to be very young, i.e. $t_{\rm Q}\lesssim 10^4-10^5$~yr \citep{Eilers2017a, Eilers2018b}. These three young quasars provide valuable clues to the accretion behavior of SMBHs, and pose significant challenges on current black hole formation models that require much longer lifetimes to explain the growth of SMBHs. 

This study aims to determine the fraction of such young objects within the quasar population at large, and to establish a statistically uniform and significant sample of young quasars, which will then enable us to search for any spectral or environmental signatures that might distinguish these young objects from the whole quasar population. 
To this end, we conduct preliminary measurements of the proximity zones $R_p$ of $122$ quasars at $5.6\lesssim z\lesssim 6.5$. However, these preliminary measurements have large uncertainties due to their imprecise redshift estimate, which constitute the largest source of uncertainty for proximity zone measurements. 

From this sample we select the best young quasar ``candidates'' whose preliminary measurements of their proximity zones are very small, and thus they potentially indicate very short quasar lifetimes. 
For these young candidates we conduct a multi-wavelength survey which we present in this paper, in order to obtain measurements of the quasars' systemic redshifts, and precisely measure the extents of their proximity zones. 
In a following paper \citep[hereafter \citetalias{E20prep}]{E20prep} we will use these measurements to derive constraints on the quasars' lifetimes, and study the dependence of spectral properties with quasar age. 

%The outline of this paper is as follows: in \S~\ref{sec:quasar_sample} we present the sample of quasars that we analyze in this study. We describe the multi-wavelength observations that we collected to characterize the quasars in \S~\ref{sec:data}. In \S~\ref{sec:analysis} we determine the systemic redshifts of the quasars, measure their proximity zones, and estimate further characteristics of the quasars, such as their emission line luminosities, star formation rates, black hole masses, and Eddington ratios. We discuss and summarize our results in \S~\ref{sec:summary}. 

Throughout this paper, we assume a flat $\Lambda$CDM cosmology of $h = 0.685$, $\Omega_m = 0.3$, and $\Omega_{\Lambda} = 0.7$, which is consistent within the $1\sigma$ errorbars with \citet{Planck2018}. 

\section{Quasar Sample}\label{sec:quasar_sample}

We target quasars that are likely to have short lifetimes, as indicated by a very small proximity zone. To this end we analyzed the spectra of $122$ quasars at $5.6\lesssim z\lesssim 6.5$, which were taken with a variety of different telescopes and instruments, and thus cover different wavelength ranges, have different spectral resolutions, and different signal-to-noise ratios \citep[see][for details]{Willott2009, Banados2016, Reed2017, Wang2017, Farina2019}. 
Based on these spectra we conducted preliminary estimates of the quasars' proximity zone sizes (see \S~\ref{sec:cont} and \S~\ref{sec:rp} for details on the procedure for measuring proximity zones). However, these measurements have large uncertainties up to $\Delta v\sim 1000\,\rm km\,s^{-1}$, i.e. $\Delta R_p\sim 1.5$~proper Mpc (pMpc), due to the highly uncertain redshift estimates, which are based on template fitting of broad rest-frame UV emission lines that can be displaced from the systemic redshift due to strong internal motions or winds. 

In order to compare the proximity zone sizes of quasars with different luminosities, we normalized these preliminary proximity zone estimates to the same absolute magnitude of $M_{1450}=-27$ \citep{Fan2006, BoltonHaehnelt2007a, Eilers2017a}. This eliminates the dependency of the zone sizes on the quasars' luminosities, since brighter quasars emit more ionizing radiation and are thus expected to have a larger proximity zone for a given quasar age. The exact procedure is described in detail in \S~\ref{sec:rp}. 
These ``corrected'' proximity zones $R_{p,\,\rm corr}$ of our sample span a range of $0.6~{\rm pMpc}~\lesssim R_{p,\,\rm corr}\lesssim 12.1$~pMpc, with a mean of $\langle R_{p,\,\rm corr}\rangle\approx 5.4$~pMpc.  

We excluded any objects that showed clear evidence for a premature truncation of the proximity zones due to associated absorption systems, such as proximate damped \lya\ absorption systems (pDLAs), which we identify by searching for ionic metal absorption lines in the quasar continuum that are associated with the absorber. Such quasars with associated absorption systems spuriously resemble quasars with small proximity zones. Additionally, we exclude quasars with clear broad absorption line (BALs) features that might contaminate the proximity zones. 

We then choose the $10\%$ of quasars from this sample with the smallest proximity zone measurements for follow--up multi--wavelength observations. These $12$ objects all exhibit proximity zones with $R_{p,\,\rm corr}< 2$~pMpc. One additional object, CFHQS\,J2229+1457, has been added to this sample for follow--up observations. This quasar was previously identified to have a small proximity zone, but the available spectrum obtained with the Low Resolution Imaging Spectrometer (LRIS) on Keck did not cover the near-infrared wavelengths, nor did it have sufficient resolution to securely exclude any premature truncation of its proximity zone \citep{Eilers2017a}. %The complete sample of $13$ quasars for this study is listed in Table~\ref{tab:overview}.} \\

\begin{deluxetable*}{llllLl}
\setlength{\tabcolsep}{9pt}
\tablecaption{Overview of our data set sorted by right ascension. \label{tab:overview}}
\tablehead{\colhead{object} & \dcolhead{\rm RA~[hms]} & \dcolhead{\rm DEC~[dms]} & \colhead{instrument} & \dcolhead{t_{\rm exp}\,\rm [hr]} & \colhead{program ID}}
\startdata
PSO\,J004+17 & 00:17:34.467	& +17:05:10.696 & ALMA & 0.6 & 2017.1.00332.S (PI: Eilers)\\
 & & & VLT/X-Shooter & 1.0 & 101.B-02720 (PI: Eilers)\\
PSO\,J011+09 & 00:45:33.566	& +09:01:56.928 & ALMA & 0.3 & 2017.1.00332.S (PI: Eilers)\\
 & & & VLT/X-Shooter & 1.0 & 101.B-02720 (PI: Eilers)\\
VDES\,J0323--4701 &03:23:40.340 & --47:11:29.400 & VLT/X-Shooter & 0.7 & 101.B-02720 (PI: Eilers)\\
VDES\,J0330--4025  & 03:30:27.920 & --40:25:16.200& VLT/X-Shooter & 0.7 & 101.B-02720 (PI: Eilers)\\
PSO\,J056--16 & 03:46:52.044	& --16:28:36.876 & ALMA & 0.3 & 2017.1.00332.S (PI: Eilers)\\
 & & & X-Shooter & 2.0 & 097.B-1070 (PI: Farina)\\
PSO\,J158--14 & 10:34:46.509	& --14:25:15.855 & ALMA & 0.3 & 2017.1.00332.S (PI: Eilers)\\
 & & & VLT/X-Shooter & 1.2 & 096.A-0418 (PI: Shanks)\\
SDSS\,1143+3808 & 11:43:38.347	&+38:08:28.823 & IRAM/NOEMA & 4.0 &  W18EF (PI: Eilers)\\
 & & & Keck/DEIMOS & 1.0 & 2017A\_U078 (PI: Hennawi)\\
 PSO\,J239--07 & 15:58:50.990 & --07:24:09.591 & ALMA & 0.4 & 2017.1.00332.S (PI: Eilers)\\
 & & & VLT/X-Shooter & 1.0 & 101.B-02720 (PI: Eilers)\\
 PSO\,J261+19 & 17:24:08.746& +19:01:43.120 & IRAM/NOEMA & 2.5 & W18EF (PI: Eilers)\\
 & & & VLT/X-Shooter & 1.0 & 101.B-02720 (PI: Eilers)\\
 PSO\,J265+41 & 17:43:43.136 & +41:24:50.191 & IRAM/NOEMA & 2.5 &  W18EF (PI: Eilers)\\
 & & & Keck/DEIMOS & 1.0 & 2017B\_U090 (PI: Hennawi)\\
CFHQS\,J2100--1715 & 21:00:54.616 & --17:15:22.500 & VLT/X-Shooter & 3.3 & 097.B-1070 (PI: Farina)\\
%PSO\,J319-10 & 21:18:24.970 & -10:55:57.430& VLT/X-Shooter & 1.0 & 101.B-02720 (PI: Eilers)\\
CFHQS\,J2229+1457  & 22:29:01.649	&+14:57:08.980& VLT/X-Shooter & 1.7 & 101.B-02720 (PI: Eilers)\\
PSO\,J359-06 & 23:56:32.451	& --06:22:59.255 & ALMA & 0.3 & 2017.1.00332.S (PI: Eilers)\\
 & & & VLT/X-Shooter & 1.3 & 098.B-0537 (PI: Farina)
 \enddata
\tablecomments{The columns show the name of the quasar, its coordinates RA and DEC in the J2000 epoch, as well as the instrument with which the data was taken, the exposure time on the source, and the program ID of the observation. }
\end{deluxetable*}

\section{Multi-wavelength Data Set}\label{sec:data}

We intend to measure the systemic redshifts by means of sub--mm observations of emission lines arising from the cold gas reservoir of the quasar host galaxies with the Atacama Large Millimetre Array (ALMA), \S~\ref{sec:alma}, and the NOrthern Extended Millimeter Array (NOEMA) at the Institute de Radioastronomie Millim{\'e}trique (IRAM), \S~\ref{sec:noema}. These emission lines provide a tenfold improvement on the quasars' systemic redshifts, because they do not suffer from possible displacements due to strong internal motions or winds in the quasars' broad line regions (BLR). 
Furthermore, we obtained deep optical (VIS) and near-infrared (NIR) spectra with the medium resolution spectrographs X-Shooter on the Very Large Telescope (VLT), \S~\ref{sec:xshooter}, as well as optical spectra for quasars located in the Northern hemisphere with the Deep Imaging Multi-Object Spectrograph (DEIMOS) on the Keck telescopes, \S~\ref{sec:Keck}, to search for associated absorption systems that might have contaminated and prematurely truncated the quasars' proximity zones. 
Table~\ref{tab:overview} shows a summary of the observations presented in this paper. 

\begin{figure*}[!ht]
\centering
\includegraphics[width=.87\textwidth]{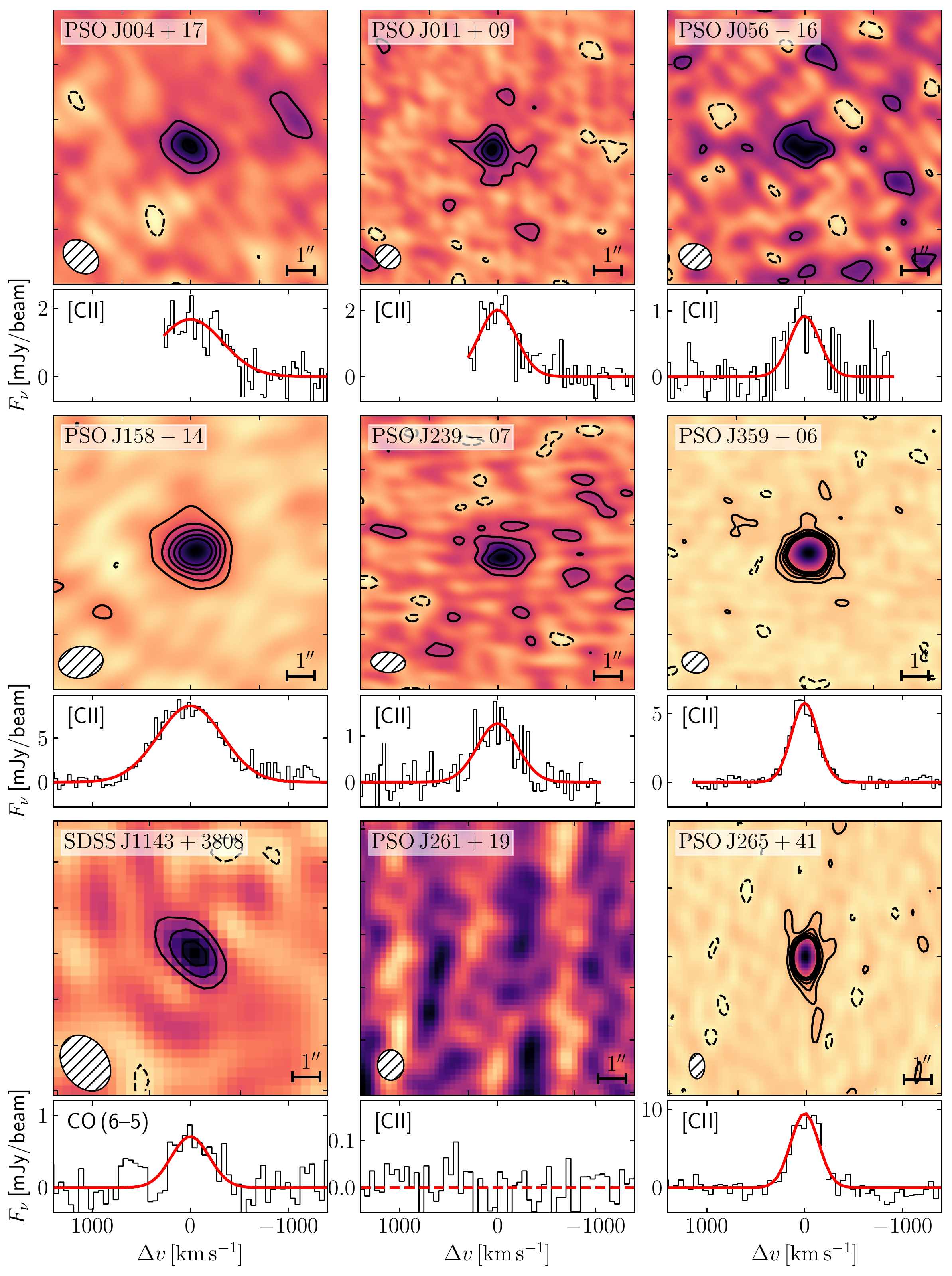}
\caption{\textbf{\cii\ and CO emission line maps.} The six quasars in the top $2$ rows show observations with ALMA, while the bottom row shows three quasars that have been observed with NOEMA. The upper panels of each source present the collapsed and continuum-subtracted line maps ($10\arcsec\times10\arcsec$ in size), where the black solid and dashed contours indicate the $\pm 2, 4, 6, 8, 10\sigma$ isophotes. The bottom panels show the extracted spectra (black) at the brightest peak position, as well as the best Gaussian fit (red) to their emission line. \label{fig:postage}} 
\end{figure*}

\subsection{ALMA Observations}\label{sec:alma}

The ALMA data consist of short ($\sim 15-20$ minutes on source) observations centered on the optical/NIR coordinates of the quasars. The tuning frequency of the spectral windows was chosen such that two neighbouring windows encompass the expected observed frequency of the \cii\ emission line at $158\,\mu$m ($\nu_{\rm rest}=1900.548$~GHz) based on preliminary redshift estimates. The other two spectral windows cover the dust continuum emission. 
Observations were carried out in May 2018 with the array in a compact configuration $\rm C43-2$, resulting in images with $\sim 1\arcsec$ spatial resolution. Thus, the size of the \cii-emitting region is comparable to the expected sizes of the quasar host galaxies and hence the sources are likely unresolved \citep{Walter2009}. 

The data were processed with the default calibration procedure making use of the CASA pipeline \citep{McMullin2007}, version $5.1.2$. The data cubes were then imaged with the CASA command \texttt{tclean} using Briggs cleaning and a robust parameter of $2$ (natural weighting), in order to maximize the signal-to-noise ratio (SNR) of our observations. The mean rms noise is $0.35\,\rm mJy\,beam^{-1}$ per $30$~MHz bin. 

The map of the continuum emission is calculated by averaging the two line-free spectral windows. We obtain the emission line map by subtracting the continuum emission applying the CASA command \texttt{uvcontsub}, and afterwards collapsing the data cube within a narrow frequency range ($\Delta\nu=0.4$~GHz for PSO\,J158--14 due to its broad \cii\ line, and $\Delta\nu=0.25$~GHz for all other objects) around the peak frequency of the emission line. All images of the collapsed and continuum-subtracted line maps are shown in the top six panels in Fig.~\ref{fig:postage}. The dust continuum maps are shown in Appendix~\ref{app:cont} in Fig.~\ref{fig:cont_map}. 

\subsection{NOEMA Observations}\label{sec:noema}

Three quasars from our sample located in the Northern hemisphere have been observed with the ten antennas of NOEMA in compact configuration (10C) in December 2018 and January 2019. Taking advantage of the PolyFix correlator we could simultaneously collect data in an upper and lower side band with a total band width of $15.4$~GHz. The data were calibrated and reduced making use of the \texttt{GILDAS} routine \texttt{clic} \citep{gildas}. 

The visibilities were imaged using the software \texttt{mapping} as part of the \texttt{GILDAS} suite. We adopt natural weighting, which results in a synthesized beam size of $0.6\arcsec\times 0.9\arcsec$ at the higher tuning frequency $\nu_{\rm obs}=270.49$~GHz for PSO\,J265+41, and $1.6\arcsec\times 2.2\arcsec$ at a lower tuning frequency of $\nu_{\rm obs}=101.14$~GHz for SDSS\,J1143+3808, for which we observed the CO(6--5) at $3\,\rm mm$ ($\nu_{\rm rest} = 691.473\rm\,GHz$) and CO(5--4) ($\nu_{\rm rest}=576.267$~GHz) emission lines instead of \cii. The imaged cubes are re-binned on a spectral axis with $50\,\rm km\,s^{-1}$ wide channels. The rms noise for SDSS\,J1143+3808 is $0.21\,\rm mJy\,beam^{-1}$ per $50\,\rm km\,s^{-1}$ bin, whereas the other two observations for PSO\,J261+19 and PSO\,J265+41 have a larger rms noise of $1.18\,\rm mJy\,beam^{-1}$ and $1.01\,\rm mJy\,beam^{-1}$ per $50\,\rm km\,s^{-1}$ bin, respectively. The continuum flux is estimated using the line-free channels, and subtracted from the cubes. The bottom three panels of Fig.~\ref{fig:postage} show the continuum subtracted line maps obtained with NOEMA. 

\subsection{VLT/X-Shooter Observations}\label{sec:xshooter}

Most quasar spectra observed with X-Shooter on the VLT were obtained on August 18th and 19th, 2018, in visitor mode (program ID: 101.B-02720). The data of four quasar spectra, i.e. PSO\,J056--16, PSO\,J158--14, CFHQS\,J2100--1715, and PSO\,J359--06 were acquired between January 2016 and May 2017, and are taken from the ESO archive\footnote{\url{http://archive.eso.org/eso/eso_archive_main.html}} \citep[][Farina et al.\ in prep.]{Chehade2018}. We obtained multiple exposures of $1200$~s each, with the $0.6\arcsec\times11\arcsec$ slit in the NIR, and the $0.9\arcsec\times11\arcsec$ slit in the VIS.
The VIS observations are binned $2\times2$ in spectral and spatial direction. Using a $0.9\arcsec$ slit width we obtain a spectral resolution of $R\approx 8900$ in the visible wavelength regime and $R\approx8100$ in the NIR arm for a $0.6\arcsec$ slit\footnote{\url{https://www.eso.org/sci/facilities/paranal/instruments/xshooter/inst.html}}.
The wavelength range covers $5500{\rm {\AA}}\lesssim \lambda_{\rm obs} \lesssim 22000{\rm {\AA}}$. We dithered the different exposures along the slit to allow for image differencing in the data reduction step (see \S~\ref{sec:Keck}).  
One object, PSO\,J158--14, has been observed with the wider $0.9\arcsec$ slit in the NIR ($R\approx5600$) and the $K$-band blocking filter, which results in a reduced wavelength coverage of $5500{\rm {\AA}}\lesssim \lambda_{\rm obs} \lesssim 20250{\rm {\AA}}$. 

\subsection{Keck/DEIMOS Observations}\label{sec:Keck}
The two quasar spectra taken with the DEIMOS instrument at the Nasmyth focus on the Keck II telescope were observed in May and September 2017. For each object we acquired three exposures of $1200$~s each. In the case of SDSS\,J1143+3808 we used a custom-made slitmask with a $1\arcsec$ slit and the $830$G grating, resulting in a pixel scale of $\Delta \lambda\approx0.47$~{\AA} and a spectral resolution of $R\approx 2500$. For PSO\,J265+41 we used the same grating, but the \texttt{LongMirr} slitmask with a narrower ($0.7\arcsec$) slit, resulting in a slightly higher resolution.
The grating was tilted to a central wavelength of $8400$~{\AA}, resulting in a wavelength coverage of $6530{\rm {\AA}}\lesssim \lambda_{\rm obs} \lesssim 10350{\rm {\AA}}$.

All optical and NIR spectroscopic data were reduced applying standard data reduction techniques with the newly developed open source python spectroscopic data reduction package \texttt{PypeIt} \citep{Prochaska2020, pypeit_zenodo}.
The reduction procedure includes sky subtraction, which was performed on the 2D images by including both image differencing between dithered exposures (whenever these were available) and a B-spline fitting procedure. In order to then extract the 1D spectra the optimal spectrum extraction technique is applied \citep{Horne1986}. The individual 1D spectra are flux calibrated using the standard stars $\rm LTT\,3218$ (for spectra observed with VLT/X-Shooter) and $\rm G191B2B$ or $\rm Feige\,110$ (for spectra taken with Keck/DEIMOS). Finally, the fluxed 1D spectra are stacked and a telluric model is fitted to the stacked spectra using telluric model grids produced from the Line-By-Line Radiative Transfer Model \citep[LBLRTM\footnote{\url{http://rtweb.aer.com/lblrtm.html}};][]{Clough2005, Gullikson2014}, resulting in the final spectra. 

\begin{figure*}[!ht]
\centering
\includegraphics[width=.95\textwidth]{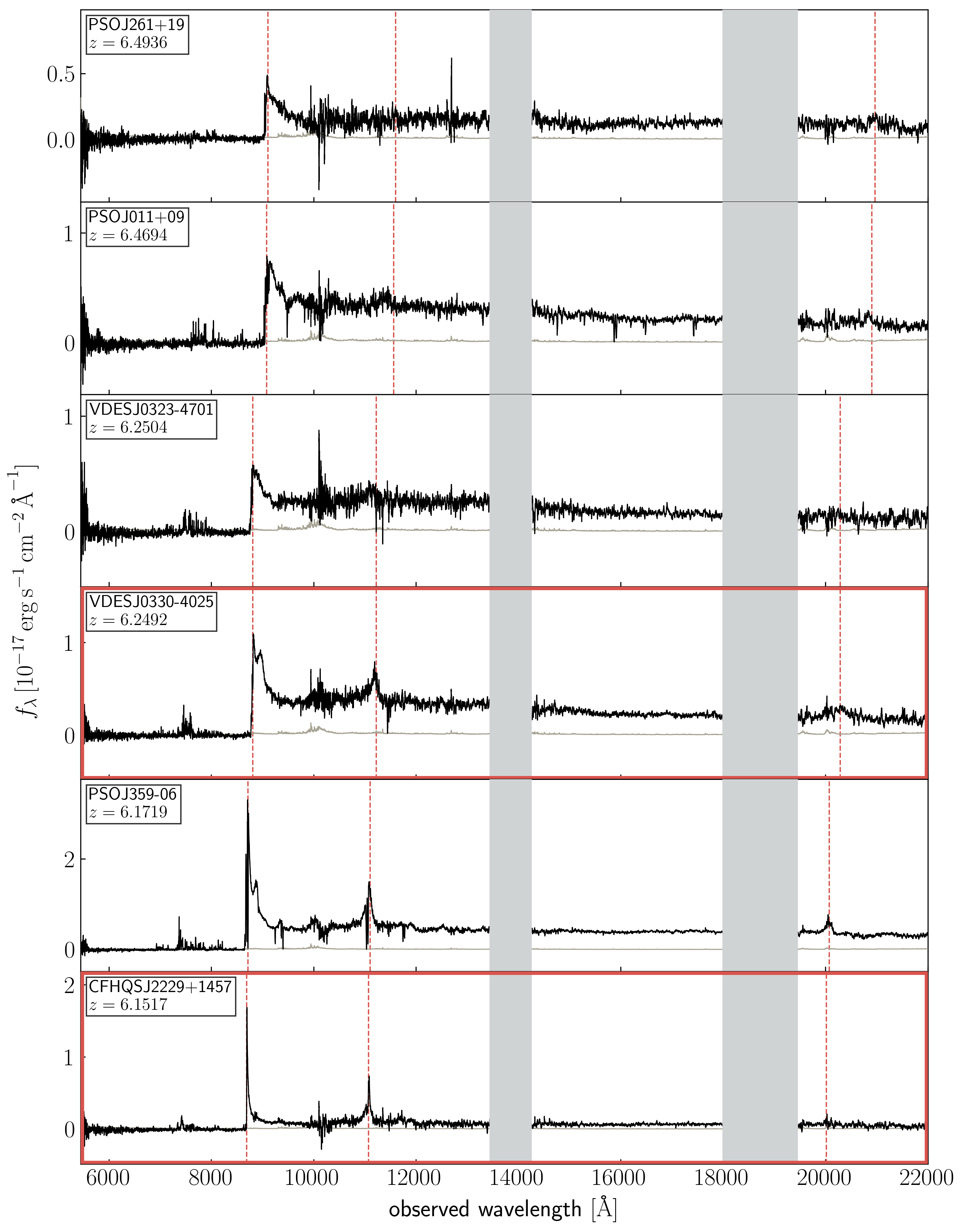}
\caption{\textbf{Spectra of the quasars in our sample.} All spectra (black) are observed with VLT/X-Shooter, besides the spectra of the two quasars PSO\,J265+41 and SDSS\,J1143+3808, which have been observed with Keck/DEIMOS. Regions of large telluric absorption have been masked with grey shaded regions. The dashed red lines indicate the location of various emission lines, i.e. \lya\ at $1215.7$~{\AA} in the rest-frame, the \ion{C}{4} doublet at $1548$~{\AA} and $1550$~{\AA}, as well as the \mgii\ emission line at $2798.7$~{\AA}. All spectra and noise vectors (grey) have been inverse-variance smoothed with a $20$ pixel filter. The red frames indicate quasars that exhibit very small proximity zones, i.e. $R_{p,\,\rm corr}\lesssim 2$~pMpc. \label{fig:spectra_a}}
\end{figure*}

\begin{figure*}[!ht]
\centering
\includegraphics[width=.95\textwidth]{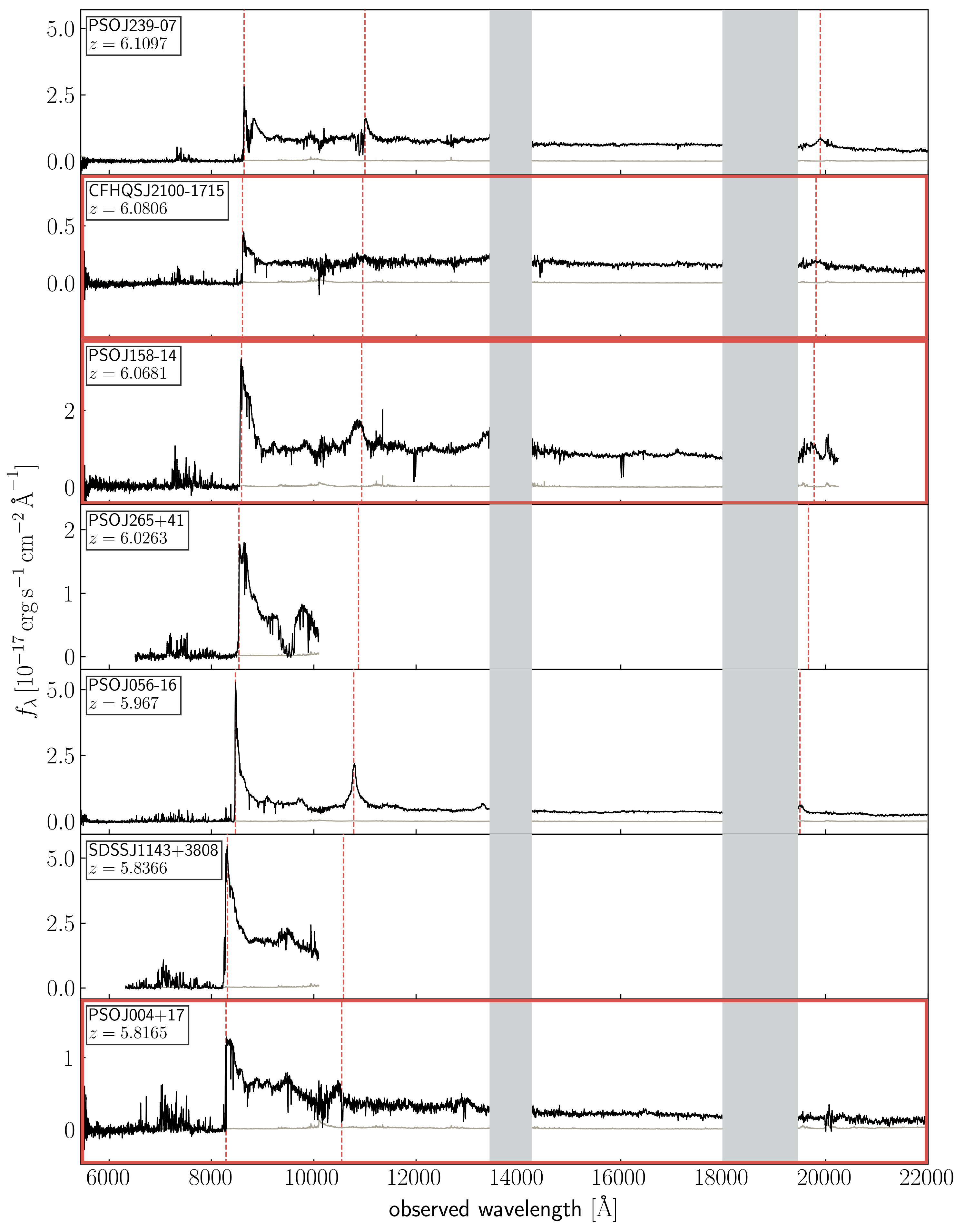}
\caption{Same as Fig~\ref{fig:spectra_a}. \label{fig:spectra_b}}
\end{figure*}

Fig.~\ref{fig:spectra_a} and Fig.~\ref{fig:spectra_b} show the final reduced optical and NIR spectra for all quasars in our sample. We apply a running $20$ pixel filter when showing the spectra and noise vectors with the average flux computed using inverse variance weights. 

\section{Analysis}\label{sec:analysis}

In this section we analyze our multi-wavelength data set to measure the quasars' systemic redshifts (\S~\ref{sec:zsys} and \S~\ref{sec:BH}), and estimate their optical continuum emission (\S~\ref{sec:cont}), in order to measure the extents of their proximity zones (\S~\ref{sec:rp}). We determine further properties of the quasars, such as the star formation rates (SFR) of the host galaxies (\S~\ref{sec:sfr}), as well as the black hole masses and the Eddington ratios of the accretion. 

\subsection{Systemic Redshifts of Quasar Host Galaxies\label{sec:zsys}}

The most precise estimates of the systemic redshifts of quasars are based on the narrow atomic or molecular emission lines arising from the gas reservoir within the quasars' host galaxy. We estimate the systemic redshifts primarily by means of the \cii\ emission line, which is the dominant coolant of the ISM. For one quasar in our sample, SDSS\,J1143+3803, this line was outside of the observable frequency range of NOEMA, and thus we observed the CO(6--5) and CO(5--4) emission lines, which usually are some of the brightest molecular CO transitions in quasars \citep[e.g.][]{CarilliWalter2013, Yang2019b}. Because these lines arise from the host galaxy itself, they provide a much more precise redshift estimate compared to rest-frame UV emission lines that arise from the BLR around quasars and may suffer from strong internal motions or winds, potentially displacing the emission line centers from the systemic redshift \citep[e.g.][]{Richards2002a, Venemans2016,  Mazzucchelli2017}. 

We extract the $1$D spectra from the continuum-subtracted data cubes at the position of the brightest emission from the source. We then fit the \cii\ or CO emission lines assuming a Gaussian line shape. We apply the Markov Chain Monte Carlo (MCMC) affine-invariant ensemble sampler \texttt{emcee} \citep{emcee} with flat priors for the amplitude $A\in[0,\,20]$~mJy, the width $\sigma\in[0,\,1]$~GHz of the emission line, as well as the peak frequency $\nu_{\rm obs}$. We adopt the median of the resulting posterior probability distribution as the best parameter estimate. We take the peak of the Gaussian fit as the best estimate for the systemic redshift of the quasar with an uncertainty arising from the $68$th percentile of the posterior probability distribution. 

The extracted spectra and the corresponding best fit to the sub-mm emission lines are shown in the lower panels of each object in Fig.~\ref{fig:postage}. Table~\ref{tab:submm2} shows the estimated parameters derived from the sub-mm emission lines of the quasars. Note that all derived flux values only include statistical uncertainties, but neglect the $\sim 10\%$ systematic uncertainty that comes from calibrating interferometric data \footnote{\url{https://almascience.nrao.edu/documents-and-tools/cycle7/alma-technical-handbook/view}}. 

Note that in one case (PSO\,J261+19) no clear detection of an emission line associated with the quasar host galaxy could be found, and we extracted the spectrum at the nominal position of the target based on optical/NIR data. Since we also do not detect any continuum emission from this source (see Fig.~\ref{fig:cont_map} in Appendix~\ref{app:continuum}), this non-detection is likely explained by a very faint emission that is below the detection limit of our $2.5$ hour exposure with NOEMA, i.e. $F_{\rm cont}<0.05\rm\,mJy$. If the non-detection of the \cii\ emission line would be explained by a wrong redshift estimate that would have shifted the emission line outside of the observable frequency range, the offset between its systemic redshift and the redshift estimate based on its \mgii\ emission line (see \S~\ref{sec:BH}) would have to be $\Delta v > 4730\,\rm km\,s^{-1}$. Such large velocity shifts have not been reported in the literature, and thus this possibility seems unlikely.

\begin{deluxetable*}{lcCCCCCC}
\setlength{\tabcolsep}{5pt}
\tablecaption{Measurements of sub-mm properties of the quasar sample. \label{tab:submm2}}
\tablehead{\colhead{object}  & \colhead{line} & \dcolhead{\nu_{\rm obs}} & \dcolhead{z_{\rm sub-mm}} & \dcolhead{\rm FWHM} & \dcolhead{F_{\rm cont}}  & \dcolhead{F_{\rm line}}& \dcolhead{\log L_{\rm line}}\\
\colhead{} & \colhead{} & \colhead{[GHz]} & \colhead{} & \dcolhead{\rm [km\,s^{-1}]} & \dcolhead{\rm [mJy]}  & \dcolhead{\rm [Jy\,km\,s^{-1}]}& \dcolhead{\rm[L_{\odot}]} }
\startdata
PSO\,J004+17 & [\ion{C}{2}] & 278.81\pm0.04 & 5.8165\pm0.0004 & 777\pm95 & 0.88\pm0.01 & 0.21\pm0.01 & 8.31\pm0.01 \\
PSO\,J011+09 & [\ion{C}{2}] & 254.44\pm0.02 & 6.4694\pm0.0002 & 449\pm66 & 1.20\pm0.01 & 0.27\pm0.01 & 8.47\pm0.01 \\
PSO\,\,J056--16 & [\ion{C}{2}] & 272.79\pm0.03 & 5.9670\pm0.0003 & 355\pm58 & 0.17\pm0.01 & 0.01\pm0.01 & 7.11\pm0.22 \\
PSO\,J158--14 & [\ion{C}{2}] & 268.89\pm0.01 & 6.0681\pm0.0001 & 780\pm27 & 3.46\pm0.02 & 1.66\pm0.02 & 9.22\pm0.01 \\
SDSS\,J1143+3808 & CO(6--5) & 101.14\pm0.01 & 5.8366\pm0.0008 & 452\pm83 & 0.05\pm0.01 & 0.72\pm0.06 & 8.39\pm0.04 \\
& CO(5--4) &84.30\pm0.01 & 5.8356\pm 0.0004 & 361\pm 54& 0.05\pm0.01 & 0.25\pm0.05& 7.86\pm0.08 \\
PSO\,J239--07 & [\ion{C}{2}] & 267.32\pm0.02 & 6.1097\pm0.0002 & 486\pm55 & 0.23\pm0.01 & 0.23\pm0.01 & 8.37\pm0.02 \\
PSO\,J261+19 & [\ion{C}{2}] & - & - & - & <0.05 & - & - \\
PSO\,J265+41 & [\ion{C}{2}] & 270.49\pm0.01 & 6.0263\pm0.0001 & 335\pm16 & 3.61\pm0.07 & 9.20\pm0.50 & 9.96\pm0.02 \\
CFHQS\,J2100--1715 & [\ion{C}{2}] & 268.39\pm0.02^b & 6.0806\pm0.0011^a & 340\pm70^a & 1.20\pm0.15^a & 1.37\pm0.14^a & 9.12\pm0.04^a \\
CFHQS\,J2229+1457 & [\ion{C}{2}] & 265.75\pm0.02^c & 6.1517\pm0.0005^c & 351\pm39^c & 0.05\pm0.03^c & 0.58\pm0.08^c & 8.78\pm0.06^c \\
PSO\,J359--06 & [\ion{C}{2}] & 265.00\pm0.01 & 6.1719\pm0.0001 & 318\pm11 & 0.68\pm0.01 & 0.45\pm0.01 & 8.67\pm0.01 \\
 \enddata
\tablecomments{The columns show the name of the quasar, the observed sub-mm emission line, its peak frequency, the derived redshift estimate, the FWHM of the observed line, the continuum and integrated line fluxes, and the line luminosity. Parameters for CFHQS\,J2100--1715 are derived by (\textit{a}) \citet{Decarli2017}, and (\textit{b}) \citet{Decarli2018}, while parameters for CFHQS\,J2229+1457 are taken from (\textit{c}) \citet{Willott2015}. }
\end{deluxetable*}

\subsection{\mgii\ Redshifts, Black Hole Masses, and Eddington Ratios}\label{sec:BH}

For the subset of quasars for which we did not obtain a sub-mm redshift measurement (VDES\,J0323-4701, VDES\,J0330-4025, and PSO\,J261+19), we estimate their redshift by means of the \mgii\ emission line at $\lambda2798.7$~{\AA} observable in the NIR spectra. 
The \mgii\ emission arises within the BLR of the quasars and may suffer from velocity shifts with respect to the systemic redshift. However, for the majority of quasars we only expect modest velocity shifts \citep{Richards2002a, Venemans2016, Mazzucchelli2017}, and thus calculate a redshift estimate $z_{\rm Mg\,II}$ from the peak of the line.
To this end, we model the quasar emission within the wavelength region around the \mgii\ emission line, i.e. $2100$~{\AA}$\leq \lambda_{\rm rest}\leq 3089$~{\AA}, as a superposition of a power-law continuum with a slope $\alpha$ arising from the quasar's accretion disk, a scaled template spectrum of the iron
lines \ion{Fe}{2} and \ion{Fe}{3}, $f_{\lambda,\,\rm iron}$, within the BLR, as well as a single Gaussian to model the \mgii\ emission line, i.e. 
\begin{equation}
  f_{\lambda} = a_0\cdot \lambda^{-\alpha} + a_1 \cdot f_{\lambda,\,\rm iron} + a_2\cdot\exp\left(-\frac{(\lambda - \mu_{\rm MgII})^2}{2\sigma^2_{\rm MgII}}\right), \label{eq:mbh_fit}
\end{equation}
where $a_0$, $a_1$, and $a_2$ denote the amplitudes of the individual components. 
We apply the iron template spectrum from \citet{VestergaardWilkes2001}, which has been derived from a narrow emission line quasar, and convolve it with a Gaussian kernel with $\rm FWHM\approx FWHM_{\rm Mg\,II}$ to mimic the quasars' broad emission lines. 

We estimate the free parameters of the fit by means of the MCMC sampler \texttt{emcee}, assuming again flat priors and adopting the median of the posterior probability distribution as our best estimate. From the peak of the \mgii\ emission line we can then derive the redshift estimate $z_{\rm MgII}$. All fits to the \mgii\ emission lines are shown in Fig.~\ref{fig:mgii} in Appendix~\ref{app:mgii}\footnote{The corner plots of the MCMC samples \citep{corner} from the \mgii\ emission line fits for the fitting parameters $a_0$, $a_1$, and $a_2$, which are the amplitudes of the individual components in Eqn.~\ref{eq:mbh_fit}, the power-law slope $\alpha$, as well as the width $\sigma_{\rm MgII}$ and mean $\mu_{\rm MgII}$ (which is estimated as a velocity offset $\Delta v$ with respect to the systemic redshift) of the emission line are available here: \url{https://doi.org/10.5281/zenodo.3997388}. }.

The bolometric luminosity is estimated based on the quasars absolute magnitudes $M_{1450}$ and the bolometric correction by \citet[][Table 3]{Runnoe2012}, which has a scatter of approximately $0.3$~dex (see their Fig.~$5$). 
In order to estimate mass of the central SMBHs we derive the monochromatic luminosity $L_{\lambda,\,3000\text{\AA}}$ from the bolometric luminosity via $L_{\rm bol} = 5.15\times 3000{\text\AA}\,L_{3000 {\text{\AA}}}$ \citep{Richards2006}, and infer the full width at half maximum (FWHM) of the \mgii\ line from the single-epoch NIR spectra. Assuming that the dynamics in the quasar's BLR are dominated by the gravitational pull of the black hole the virial theorem can be applied, and thus we estimate the mass of the black hole by means of the scaling relation
\begin{equation}
    \frac{M_{\rm BH}}{M_{\odot}}=10^{6.86}\left(\frac{\rm FWHM_{\rm MgII}}{10^3\,\rm km\,s^{-1}}\right)^2\left(\frac{\lambda L_{\lambda,\,3000\text{\AA}}}{10^{44}\,\rm erg\,s^{-1}}\right)^{0.5}, \nonumber
\end{equation}
which has been calibrated using scaling relations from other emission lines with several thousand quasar spectra from the Sloan Digital Sky Survey \citep[SDSS;][]{VestergaardOsmer2009}. This scaling relation has an intrinsic scatter of approximately $0.55\rm\,dex$. 

Knowing the black hole masses we can derive the Eddington luminosity $L_{\rm Edd}$ of the quasars, as well as the Eddington ratio of their accretion, i.e. $\lambda_{\rm Edd} = L_{\rm bol}/L_{\rm Edd}$. 
All measurements of the NIR properties are shown in Table~\ref{tab:nir}.

\subsection{Quasar Continuum Estimates}\label{sec:cont}

Measurements of proximity zone sizes require a prediction for the underlying quasar continua. We estimate the quasar continua by means of a principal component analysis (PCA) that decomposes a set of training spectra into an orthogonal basis \citep{Suzuki2005, Paris2011, Davies2018}. Following \citet{Davies2018}, we construct a PCA decomposition based on $12,764$ training spectra from the SDSS BOSS sample in the logarithmic flux space, such that each quasar spectrum $f_{\lambda}$ can be approximated via 
\begin{equation}
    \log f_{\lambda} \approx \langle \log f_{\lambda}\rangle + \sum_{i=0}^{N_{\rm PCA}}a_i A_i, 
\end{equation}
where $\langle \log f_{\lambda}\rangle$ is the mean logarithmic flux and $A_i$ are the PCA components, weighted by the coefficients $a_i$. The logarithmic space has been chosen, because variations of the power-law quasar continuum are more naturally described by a multiplicative component rather than additive components \citep[e.g.][]{Lee2012}. 

\begin{deluxetable*}{lCCCCCC}[!t]
\setlength{\tabcolsep}{8pt}
\tablecaption{NIR properties of our quasar sample. \label{tab:nir}}
\tablehead{\colhead{object} & \dcolhead{z_{\rm Mg\,II}} & \dcolhead{\rm FWHM_{\rm Mg\,II}} & \colhead{$\Delta v$\,(\ion{Mg}{2}-[\ion{C}{2}])} & \dcolhead{M_{\rm BH}} & \dcolhead{\log L_{\rm bol}} & \dcolhead{\lambda_{\rm Edd}}\\
\colhead{} & \colhead{} & \dcolhead{\rm [km\,s^{-1}]} & \dcolhead{\rm [km\,s^{-1}]} & \dcolhead{\rm [10^9\,M_{\odot}]} & \dcolhead{\rm [erg\,s^{-1}]} & \dcolhead{}}\tablecolumns{7}
\startdata
PSO\,J011+09 & 6.444\pm0.004 & 3477\pm586 & -1021\pm143 & 1.39\pm0.47 & 47.11 & 0.72\pm0.24 \\
VDES\,J0323--4701 & 6.241\pm0.002 & 1862\pm654 & - & 0.28\pm0.20 & 46.81 & 1.76\pm1.24 \\
VDES\,J0330--4025 & 6.239\pm0.004 & 7197\pm360 & - & 4.96\pm0.51 & 46.95 & 0.14\pm0.01 \\
PSO\,J056--16 & 5.975\pm0.001 & 2556\pm79 & 339\pm27 & 0.71\pm0.04 & 47.06 & 1.26\pm0.08 \\
PSO\,J158--14 & 6.052\pm0.001 & 3286\pm127 & -673\pm49 & 1.57\pm0.12 & 47.31 & 1.01\pm0.08 \\
PSO\,J239--07 & 6.114\pm0.001 & 4490\pm64 & 195\pm25 & 2.99\pm0.09 & 47.33 & 0.55\pm0.02 \\
PSO\,J261+19 & 6.484\pm0.002 & 2587\pm183 & - & 0.47\pm0.07 & 46.69 & 0.80\pm0.12 \\
CFHQS\,J2100--1715 & 6.082\pm0.002 & 5720\pm277 & 47\pm80 & 2.18\pm0.21 & 46.64 & 0.15\pm0.02 \\
CFHQS\,J2229+1457 & 6.144\pm0.006 & 5469\pm439 & -321\pm234 & 1.44\pm0.25 & 46.36 & 0.12\pm0.02 \\
PSO\,J359--06 & 6.164\pm0.001 & 3071\pm88 & -319\pm29 & 1.05\pm0.06 & 47.09 & 0.90\pm0.05 \\
\enddata
\tablecomments{The columns show the name of the quasar, its redshift estimate based on the \mgii\ emission line, the FWHM of the \mgii\ line, the velocity shift between \mgii\ and \cii\ emission lines, the mass of the central SMBH, as well as the quasar's bolometric luminosity, and Eddington ratio. }
\end{deluxetable*}

Because quasar spectra at high redshifts suffer from significant absorption bluewards of the \lya\ emission line due to residual neutral hydrogen in the IGM, we follow the approach of previous work \citep{Suzuki2005, Paris2011}, and estimate the PCA coefficients only on the red side of the spectra. 
To this end, we construct a set $10$ ``red'' PCA components $R_i$ between $1220$~{\AA}~$\leq \lambda_{\rm rest}\leq2850$~{\AA}, as well as a set of $6$ ``blue'' PCA components $B_j$ between $1175$~{\AA}~$\leq \lambda_{\rm rest}<1220$~{\AA}. We then determine the best estimate for the coefficients for the set of red PCA components $r_i$ by fitting them to the red side of the quasar spectra, which we first normalize to unity at $\lambda_{\rm rest} = (1290\pm2.5)$~{\AA}. All spectral regions that show contamination by BAL features are masked when estimating the quasar continua. 
Note that whenever we have no NIR data available the wavelength range is truncated to $1220$~{\AA}~$\leq \lambda_{\rm rest}\leq1470$~{\AA}.  

The best set of estimated red coefficients $r_i$ are then projected onto a set of blue coefficients $b_j$ for the blue PCA components by means of a projection matrix $P_{ij}$ determined from the training spectra, i.e. 
\begin{equation}
    b_j = \sum_{i=1}^{N_{{\rm PCA},r}}r_i P_{ij}. 
\end{equation}

The quasar spectra as well as their best estimated continuum model for both the red and blue wavelength side are shown in Fig.~\ref{fig:spectra_cont_a} and Fig.~\ref{fig:spectra_cont_b} in Appendix~\ref{app:cont}. The predicted continua match the data overall well. However, while estimates of the IGM neutral gas fraction for which this continuum fitting machinery was originally developed \citep{Davies2018} critically depend on precise continuum estimates, the proximity zone measurements are more robust with respect to uncertainties in the continuum fit \citep{Eilers2017a, Eilers2017b}. The continuum model is predicted to be biased by less than $1\%$ with continuum uncertainties of less than $10\%$ in the wavelength range of interest \citep[see][Fig.~9]{Davies2018}, which only influences the proximity zone size measurement very mildly by $\langle\Delta R_p\rangle\approx 0.02$~pMpc on average. More details on the continuum uncertainties and their influence on proximity zone measurements can be found in Appendix~\ref{app:continuum_uncertainties}.

\subsection{Proximity Zone Sizes}\label{sec:rp}

\begin{figure*}[!h]
\centering
\includegraphics[width=.9\textwidth]{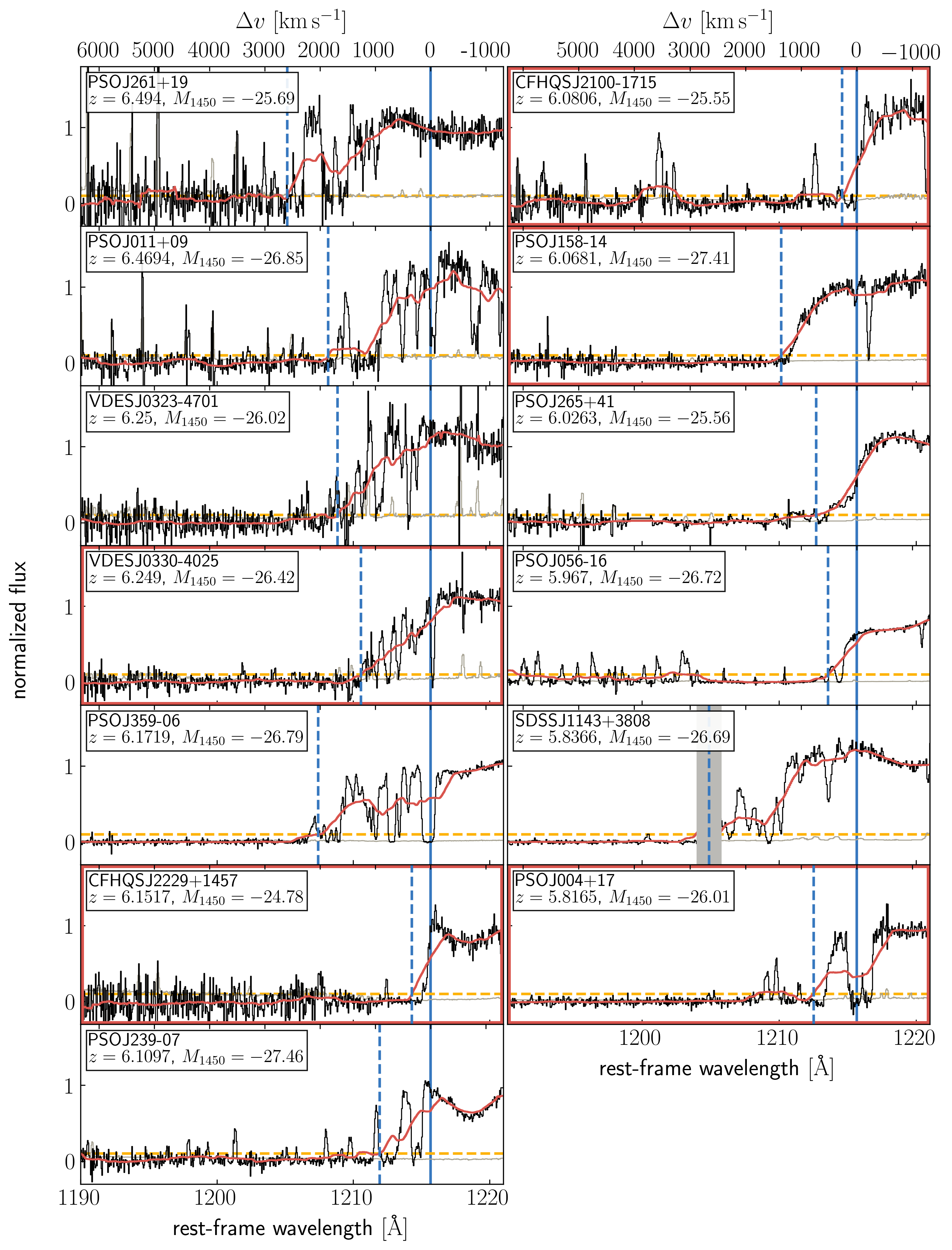}
\caption{\textbf{Proximity zones of all quasars in our sample.} The continuum-normalized fluxes with uncertainties are shown in black and grey, respectively, and are inverse-variance smoothed with a $2$ pixel filter. The blue solid and dashed lines indicate the systemic redshift of the quasar, as well as the edge of the proximity zone, respectively, which is defined to end where the smoothed continuum-normalized flux (red curves) drops below the $10\%$ flux level indicated by the yellow dashed line. The red frames show quasars with very small proximity zones. The DEIMOS detector gap in the spectrum of SDSS\,1143+3803 is masked by the grey shaded region. \label{fig:rp}} 
\end{figure*}

In order to estimate the sizes of the quasars' proximity zones, we adopt the standard definition applied in previous studies \citep{Fan2006, Willott2007, Willott2010, Carilli2010, Venemans2015, Mazzucchelli2017, Eilers2017a}. Namely, we normalize the quasar spectra by their estimated continuum emission, and smooth the continuum normalized flux with a $20$~{\AA}-wide (in the observed wavelength frame) boxcar function, which corresponds to a smoothing scale of approximately $1$~pMpc or $700\,\rm km\,s^{-1}$ at $z\sim 6$. The location at which the smoothed continuum-normalized flux drops below the $10\%$ level marks the extent of the proximity zone $R_p$. All continuum-normalized quasar spectra and their proximity zones are shown in Fig.~\ref{fig:rp}. 

For the estimate of the proximity zone sizes we take the best available redshift estimate (see \S~\ref{sec:zsys} and \S~\ref{sec:BH}). 
Note that the redshift estimates based on the \mgii\ emission line have a systematic blueshift and a systematic uncertainty compared to the systemic redshift estimate based on sub-mm emission lines (Schindler et al.\ in prep.), i.e. 
\begin{equation}
 \Delta v_{\rm MgII-[CII]} =-390^{+460}_{-280}\,\rm km\,s^{-1}, \label{eq:vshift}  
\end{equation}
which significantly dominates over the statistical uncertainty from the Gaussian fit to the peak of the \mgii\ emission line. Thus, in order to obtain a better estimate of the systemic redshift for the objects without sub-mm observations, we shift the \mgii\ redshift estimates and adapt the systematic uncertainty of these emission line shifts according to Eqn.~\ref{eq:vshift}, which is the dominant source of uncertainty on the proximity zone measurements. We also account conservatively for a systematic uncertainty of $\Delta v=100\,\rm km\,s^{-1}$ on the systemic redshifts based on sub--mm estimates, which corresponds to $\Delta z\approx 0.0024$ at $z\approx 6$. 

The size of the proximity zone also depends on the luminosity of quasars \citep[e.g.][]{Fan2006, BoltonHaehnelt2007a, Davies2019a}, since more luminous quasars emit more ionizing radiation at any given quasar age, and thus we normalize the proximity zone measurements to the same fiducial absolute luminosity of $M_{1450}=-27$. 
Theoretically, the dependency between the proximity zones and the rate of ionizing photons emitted by the quasar $\dot{N}_{\gamma}$ is $R_p\propto \dot{N}_{\gamma}^{1/2}$ \citep{BoltonHaehnelt2007a, Davies2019a}. However, due to additional heating from the reionization of \ion{He}{2} within the proximity zone this relation might differ in practice. Using a radiative transfer simulation we found that a relation between the observed proximity zone sizes and quasar luminosity of 
\begin{align}
    R_{p,\,\rm corr} = R_p \times 10^{-0.4(-27-M_{1450})/2.35}\label{eq:correction}
\end{align}
best eliminates the dependency on the quasars' luminosity and yields ``luminosity-corrected'' proximity zone measurements $R_{p,\,\rm corr}$. 
Note that this scaling relation depends only marginally on the ionization state of the ambient IGM surrounding the quasars \citep{Eilers2017a}. 
All proximity zone measurements, as well as the luminosity-corrected estimates, are presented in Table~\ref{tab:rp}.  

\begin{deluxetable*}{lCcLCCc}[!t]
\setlength{\tabcolsep}{12pt}
\tablecaption{Proximity zone measurements. \label{tab:rp}}
\tablehead{\colhead{object} & \dcolhead{z (\pm \sigma_{\rm sys})} & \dcolhead{z_{\rm line}}& \dcolhead{M_{1450}} & \dcolhead{R_p} & \dcolhead{R_{p,\,\rm corr}} & \colhead{notes}\\
\colhead{} & \colhead{} & \colhead{} & \colhead{} & \colhead{[pMpc]} & \colhead{[pMpc]} & \dcolhead{}}
\startdata
PSO\,J004+17 & 5.8165\pm0.0023 & [\ion{C}{2}] & -26.01 & 1.16\pm0.15 & 1.71\pm0.22 & -- \\
PSO\,J011+09 & 6.4694\pm0.0025 & [\ion{C}{2}] & -26.85 & 2.40\pm0.13 & 2.55\pm0.14 & -- \\
VDES\,J0323--4701 & 6.249^{+0.011}_{-0.007} & \ion{Mg}{2} & -26.02 & 2.27^{+0.62}_{-0.38} & 3.33^{+0.91}_{-0.56} & -- \\
VDES\,J0330--4025 & 6.249^{+0.011}_{-0.007} & \ion{Mg}{2} & -26.42 & 1.68^{+0.62}_{-0.38} & 2.11^{+0.78}_{-0.48} & -- \\
PSO\,J056--16 & 5.9670\pm0.0023 & [\ion{C}{2}] & -26.72 & 0.75\pm0.14 & 0.83\pm0.16 & pDLA \\
PSO\,J158--14 & 6.0681\pm0.0024 & [\ion{C}{2}] & -27.41 & 1.91\pm0.14 & 1.63\pm0.12 & -- \\
SDSS\,J1143+3808 & 5.8366\pm0.0023 & CO(6--5) & -26.69 & 3.93\pm0.63 & 4.44\pm0.71 & -- \\
PSO\,J239--07 & 6.1097\pm0.0024 & [\ion{C}{2}] & -27.46 & 1.29\pm0.14 & 1.07\pm0.12 & BAL \\
PSO\,J261+19 & 6.494^{+0.011}_{-0.007} & \ion{Mg}{2} & -25.69 & 3.35^{+0.59}_{-0.36} & 5.60^{+0.99}_{-0.60} & -- \\
PSO\,J265+41 & 6.0263\pm0.0023 & [\ion{C}{2}] & -25.56 & 1.04\pm0.14 & 1.83\pm0.25 & BAL \\
CFHQS\,J2100--1715 & 6.0806\pm0.0024 & [\ion{C}{2}] & -25.55 & 0.37\pm0.14 & 0.66\pm0.25 & -- \\
CFHQS\,J2229+1457 & 6.1517\pm0.0024 & [\ion{C}{2}] & -24.78 & 0.47\pm0.14 & 1.12\pm0.33 & -- \\
PSO\,J359--06 & 6.1719\pm0.0024 & [\ion{C}{2}] & -26.79 & 2.80\pm0.14 & 3.04\pm0.15 & -- \\
 \enddata
\tablecomments{The columns show the name of the quasar, its best systemic redshift estimate with its systematic uncertainty, the emission line it is derived from, its absolute magnitude $M_{1450}$, as well as the size of the proximity zone and its magnitude corrected value. The last column indicates whether the quasar has broad absorption lines (BAL) or associated absorption systems, which might have contaminated the proximity zones. }
\end{deluxetable*}

\subsubsection{Search for Associated Absorption Systems}\label{sec:absorption}

We carefully search for any associated absorption systems in the quasar spectra that might prematurely truncate the quasars' proximity zones. This would be the case if a self-shielding Lyman limit system (LLS) is located within $\lesssim 1000\,\rm km\,s^{-1}$ of the quasar, around the edge of its proximity zone \citep[e.g.][]{Dodorico2018, Banados2019}. Thus we search for strong low-ionization metal absorption lines redwards of the \lya emission line, which we would expect to find if a self-shielding absorption system is present. 

\begin{figure*}[!t]
\centering
\includegraphics[width=\textwidth]{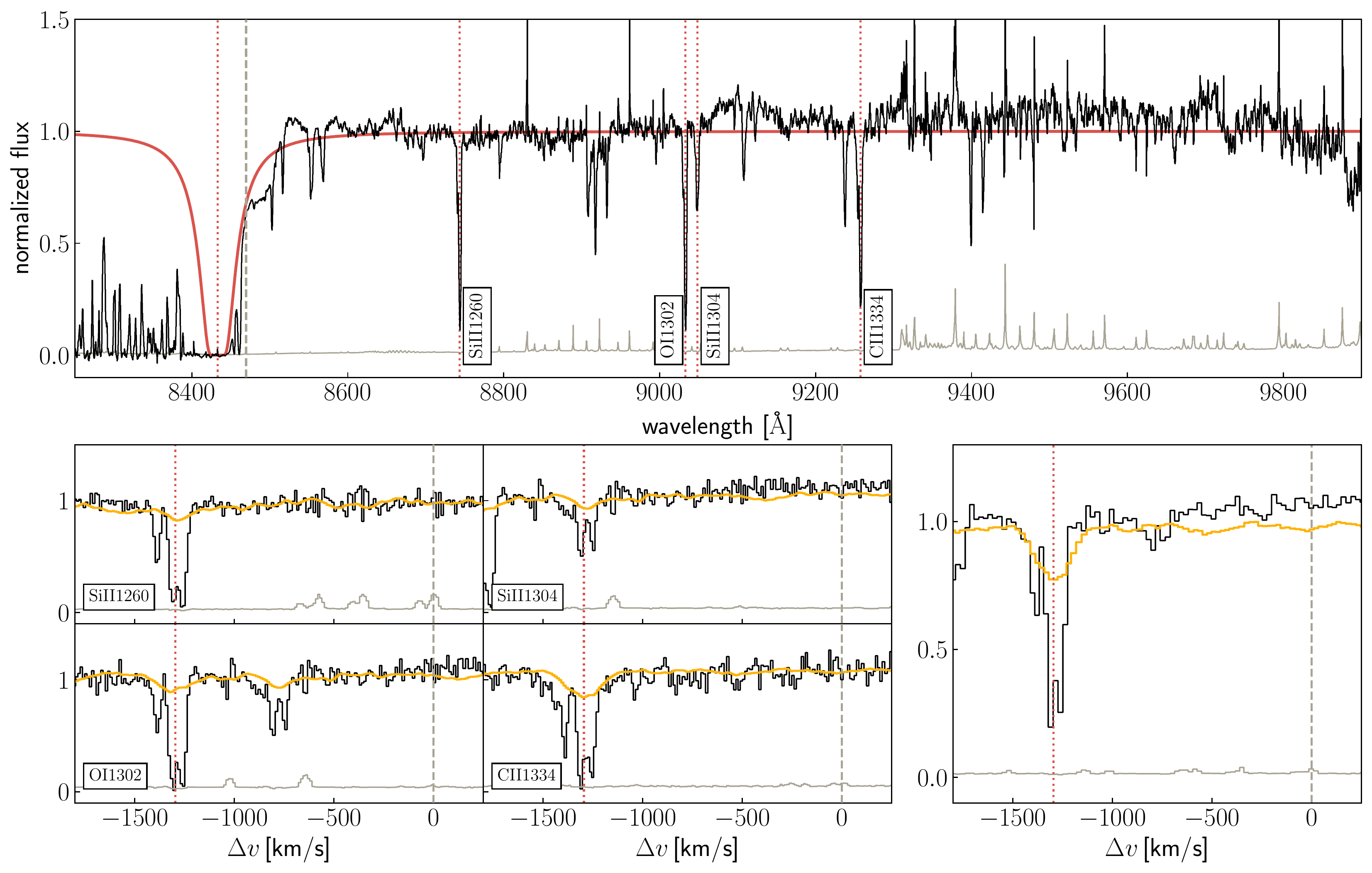}
\caption{Optical quasar spectrum (black) showing a proximate absorption systems (red) in the vicinity of PSO\,J056--16 (\textit{top}). The parameters of the absorption system, i.e. $\log N_{\rm HI}=20.4\,\rm cm^{-2}$ and $b=25\,\rm km\, s^{-1}$, are only fitted by eye. The spectrum at the locations of the low--ionization metal lines (\textit{bottom left}) and the stack thereof (\textit{bottom right}) show clear evidence for the presence an optically thick, self-shielding system. The yellow line represent a composite spectrum of 20 LLSs \citep{Fumagalli2011}. \label{fig:absorption}} 
\end{figure*}

\begin{figure*}[!t]
\centering
\includegraphics[width=\textwidth]{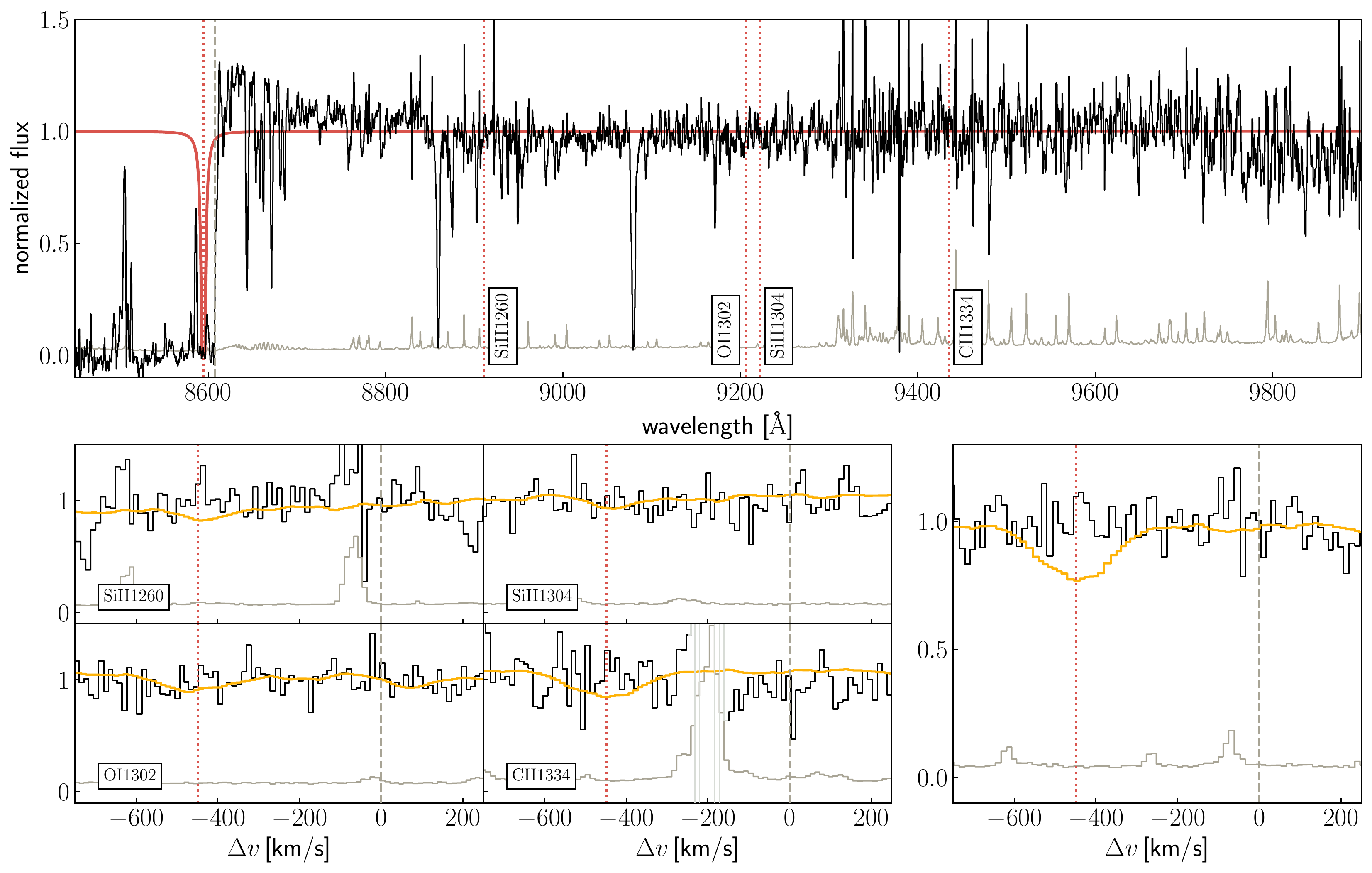}
\caption{Optical quasar spectrum (black) showing a hypothetical absorption system (red) with parameters $\log N_{\rm HI}=17\,\rm cm^{-2}$ and $b=25\,\rm km\, s^{-1}$ in the vicinity of CFHQS\,J2100--1715 (\textit{top}). The spectrum at the locations of the low--ionization metal lines (\textit{bottom left}) and the stack thereof (\textit{bottom right}) do not show evidence for the presence of such a system. The yellow line represent a composite spectrum of 20 LLSs \citep{Fumagalli2011}. Wavelength regions around very prominent skylines (shown in light grey) have been masked to avoid biases in the stack. \label{fig:absorption2}} 
\end{figure*}

To this end, we place a hypothetical absorption system at the end of each quasar's proximity zone and stack the spectrum at the location, where low-ionization metal absorption lines, i.e. \ion{Si}{2} $\lambda1260$, \ion{Si}{2} $\lambda1304$, \ion{O}{1} $\lambda1302$ and \ion{C}{2} $\lambda1334$, would fall. We compare the stacked spectrum to a composite spectrum of 20 LLSs by \citet{Fumagalli2011} in Fig.~\ref{fig:absorption} and \ref{fig:absorption2}. 

The spectrum of PSO\,J056--16 (Fig.~\ref{fig:absorption}) shows a proximate damped \lya\ absorption (pDLA) system in front of the quasar along our line-of-sight at $z_{\rm abs}\approx5.9369$, i.e. with a velocity offset of $\Delta v \approx 1297\,\rm km\,s^{-1}$. This system clearly shows low-ionization absorption lines, it has a high column density (a fit by eye indicates $N_{\rm HI}\gtrsim 10^{20}\,\rm cm^{-2}$) and is thus optically thick, causing a premature truncation of the quasar's proximity zone. We searched the dust-continuum map of this quasar shown in Fig.~\ref{fig:cont_map} in Appendix~\ref{app:continuum} for the presence of a second continuum source that could be associated with the pDLA, but we did not detect any other sources in the vicinity of this quasar, presumably due to their low continuum luminosity. 

We do not find evidence for proximate self-shielding absorption systems truncating the proximity zones in the remaining quasar spectra. Fig.~\ref{fig:absorption2} shows a hypothetical absorption system in the spectrum of CFHQS\,J2100--1715. The stacked spectrum at the location where the low-ionization metal absorption lines would fall, does not reveal any evidence for the presence of such an absorption system. The same figures for all remaining quasars with small proximity zones are shown in Appendix \ref{app:abs}. 

We exclude the quasar PSO\,J056--16 from any further analysis of its proximity zone due to its pDLA. Furthermore, two quasars in our sample, i.e. PSO\,J239--07 and PSO\,J265+41, will be excluded from any further analyses of their proximity zones, since they exhibit BAL features in their optical/NIR spectra (see Fig.~\ref{fig:spectra_b}), which might contaminate or prematurely truncate their proximity zones. 

\begin{figure*}[!t]
\centering
\includegraphics[width=\textwidth]{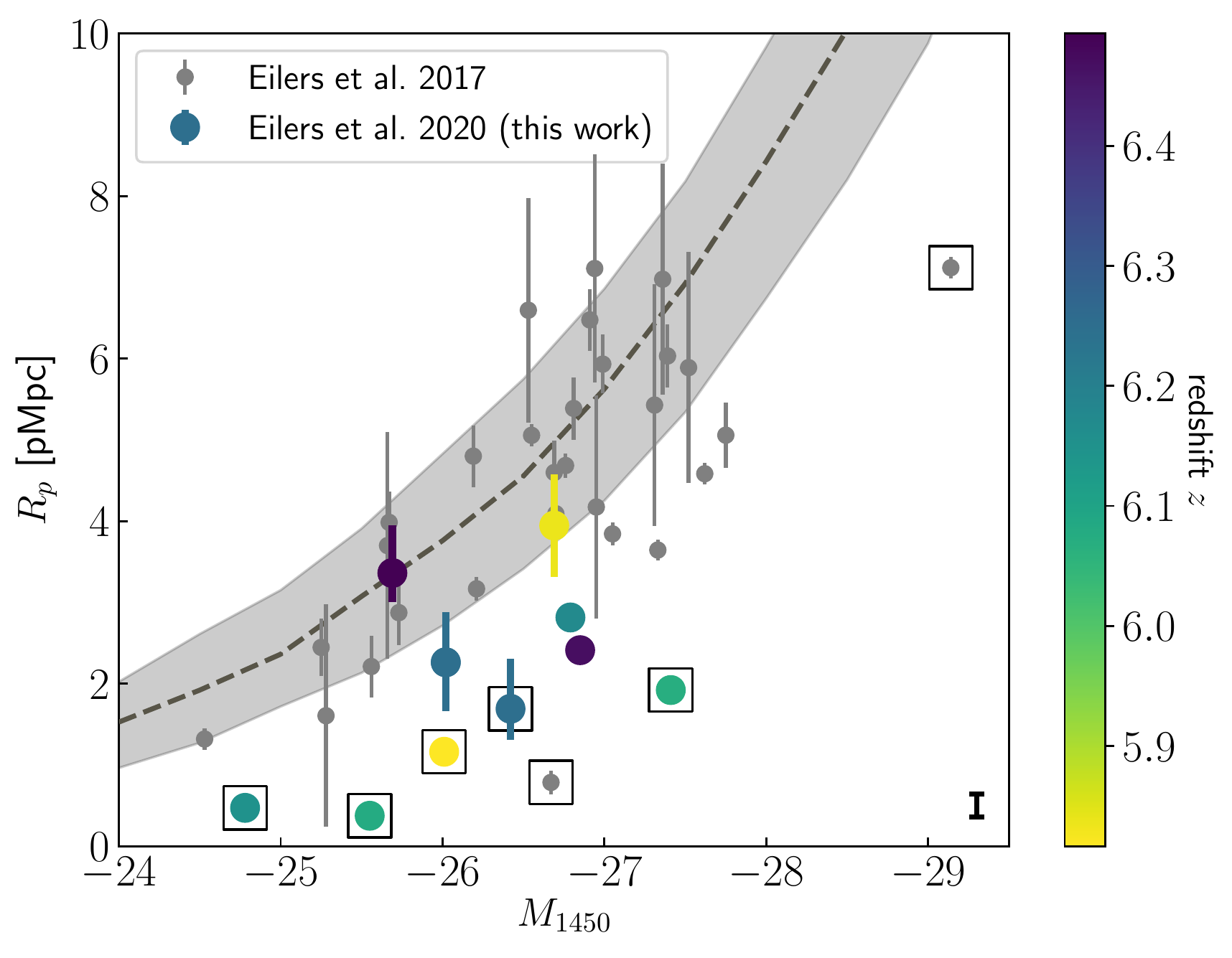}
\caption{Proximity zone sizes $R_p$ as a function of the quasars' absolute magnitude $M_{1450}$. Our new measurements are shown as colored data points, whereas the grey square data points shows previous work \citep{Eilers2017a}. The grey dashed line and shaded region represents the median and $68$th percentile of the distribution of $400$ simulated proximity zones from radiative transfer simulations at $z=6$ and $\log t_{\rm Q}=7.5$~yr \citep[see][for details]{Davies2016, Eilers2017a}, respectively. The seven quasars with extremely small proximity zones are indicated by the black boxes. The black errorbar in the bottom right indicates the systematic uncertainty on the $R_p$ measurements with sub-mm redshifts, assuming a systematic uncertainty on the sub-mm redshift estimates of $\Delta v=100\,\rm km\,s^{-1}$. \label{fig:RpM}} 
\end{figure*}

\begin{figure}[!t]
\centering
\includegraphics[width=.5\textwidth]{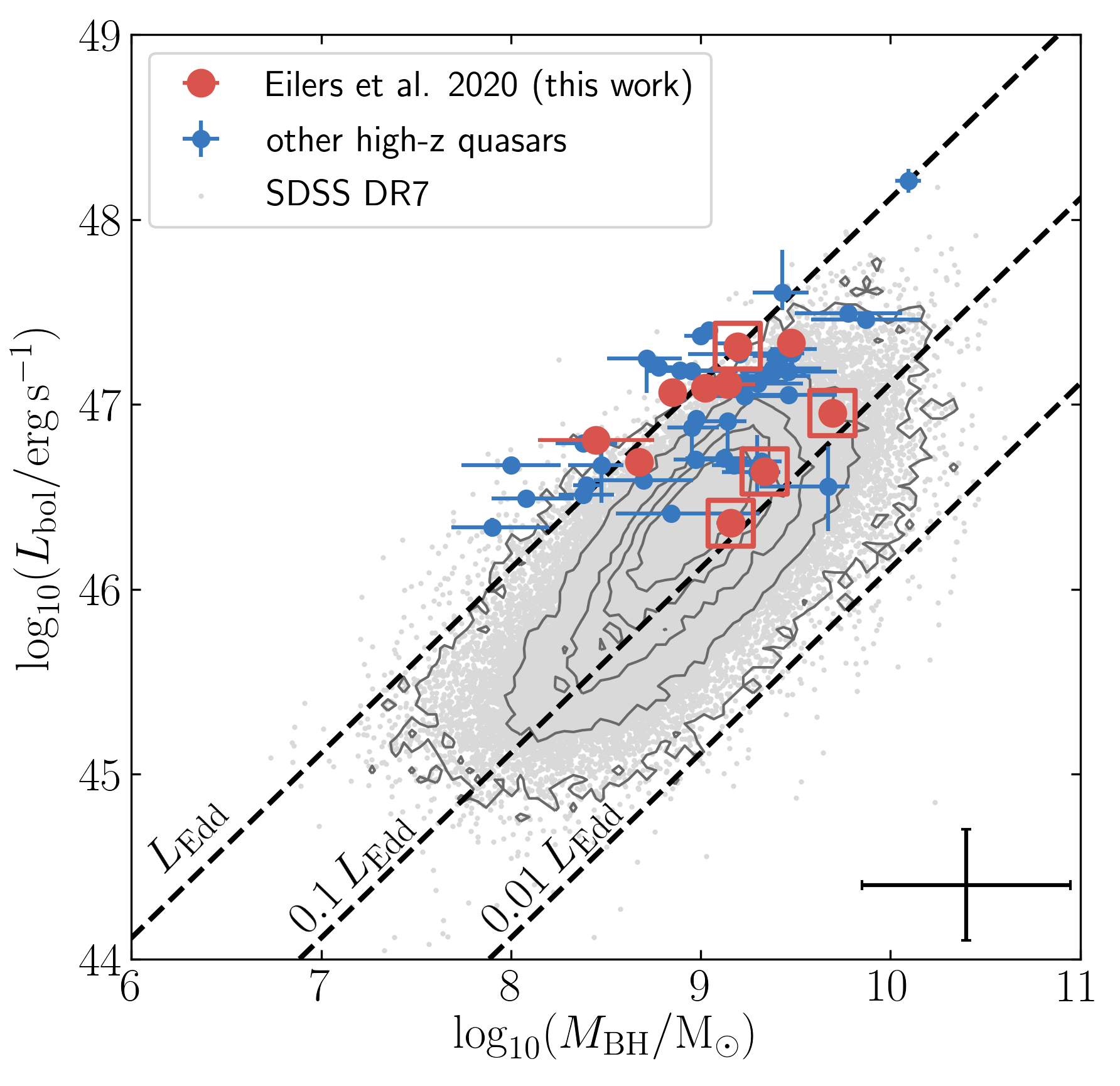}
\caption{Relation between $L_{\rm bol}$ and $M_{\rm BH}$ for our new measurements (red) compared to a quasar sample at $0.4\lesssim z\lesssim 2.2$ from SDSS DR7 (grey), as well as several other $z\gtrsim 5.8$ quasars (blue). The red boxes indicate quasars with short lifetimes. The black dashed curves show regions with constant Eddington luminosity. The errorbar on the lower right corner indicates the size of the systematic uncertainties of approximately $0.3$~dex and $0.55$~dex on $L_{\rm bol}$ and $M_{\rm BH}$, respectively. \label{fig:LM}} 
\end{figure}

\subsection{Star Formation Rates}\label{sec:sfr}

An estimate of the SFR within the quasars' host galaxies can be obtained by means of the total line fluxes $F_{\rm line}$ of the sub-mm emission lines, as well as by via the dust continuum flux $F_{\rm cont}$. To calculate these fluxes we sum the flux of all pixels within a radius of $2''$ around the source in the emission line map and the continuum map, respectively. 
We then convert the integrated line fluxes $F_{\rm line}$ into line luminosities by means of 
\begin{equation}
\frac{L_{\rm line}}{L_{\sun}} = 1.04 \times 10^{-3}\,\frac{F_{\rm line}}{{\rm Jy\,km\,s^{-1}}}\frac{\nu_{\rm obs}}{{\rm GHz}}\left(\frac{D_{\rm L}}{{\rm Mpc}}\right)^2,  
\end{equation}
where $D_{\rm L}$ represents the luminosity distance and $\nu_{\rm obs}$ is the mean observed frequency of the line \citep[see e.g.][]{CarilliWalter2013}. 

Following \citet{DeLooze2014} we estimate the SFR within the quasars' host galaxies based on their \cii\ line luminosities via 
\begin{equation}
\log \frac{\rm SFR_{\rm [C\,II]}}{M_{\odot}\,\rm yr^{-1}}= -8.52 + 1.18 \times \log \frac{L_{\rm [C\,II]}}{L_{\odot}}. \label{eq:sfr}
%\frac{\rm SFR}{\rm M_{\sun}\,yr^{-1}} = 0.052 \left(\frac{L_{\rm line}\,\Psi(y)}{10^{40}\rm\, erg\,s^{-1}}\right)^{1.034}
\end{equation}
This relation has been derived based on $z>0.5$ galaxies and has an estimated scatter of $0.4$~dex. 

An alternative method to estimate the SFR within the quasars' host galaxies is based on the dust continuum. To this end, we estimate the dust continuum emission as a modified black body \citep[e.g.][]{Dunne2000, Beelen2006}, i.e. 
\begin{equation}
    L_{\nu,\,\rm dust} = \frac{2h\nu^3}{c^2}\kappa_{\nu}(\beta)\frac{M_{\rm dust}}{\exp[h\nu/k_{\rm b}T_{\rm dust}]-1}\label{eq:Ldust}
\end{equation}
with a dust temperature of $T_{\rm dust}=47\pm3\,\rm K$, and the opacity law $\kappa_{\nu}(\beta) = 0.77(\nu/352\,{\rm GHz})^{\beta}\rm\,cm^{2}\,g^{-1}$ with the (dust) emissivity index $\beta=1.6\pm0.1$. 
We obtain an estimate of the dust mass by means of the dust continuum flux $F_{\rm cont}$, i.e. 
\begin{equation}
    M_{\rm dust} = \frac{F_{\rm cont}\,D_{\rm L}^2}{(1+z)\kappa_{\nu}(\beta) B(\nu, T_{\rm dust})}, 
\end{equation}
where $B(\nu, T_{\rm dust})$ is the Planck function and $\nu$ the rest-frame frequency \citep{Venemans2012}. The IR-luminosity $L_{\rm IR}$ is estimated by integrating Eqn.~\ref{eq:Ldust} over the solid angle and between $3\,\mu\rm m$ and $1100\,\mu\rm m$ in the rest-frame \citep[e.g.][]{KennicuttEvans2012}. 

We obtain a dust-based SFR estimate following \citet{KennicuttEvans2012} via the scaling relation
\begin{equation}
\frac{\rm SFR_{\rm dust}}{\rm M_{\sun}\,yr^{-1}} = 1.49 \times 10^{-10} \frac{L_{\rm IR}}{L_{\sun}}. \label{eq:dust_sfr}
\end{equation}
All derived quantities are shown in Table~\ref{tab:submm}. Note that the statistical uncertainties on the SFR estimates are small, but in practice the errors are dominated by systematic uncertainties arising from the scatter of $0.4-0.5$~dex in the scaling relations (Eqn.~\ref{eq:sfr} and Eqn.~\ref{eq:dust_sfr}), as well as assumptions about the dust temperature and the emissivity index, which can influence the derived SFRs by a factor of $2-3$. For more details on the systematic uncertainties on these measurements we refer the reader to section $\S 4.1$ in \citet{Venemans2018}. 

\begin{deluxetable*}{lCCCC}
\setlength{\tabcolsep}{20 pt}
\tablecaption{Properties of the quasar sample derived from the dust continuum and \cii\ emission. \label{tab:submm}}
\tablehead{\colhead{object} & \dcolhead{\log L_{\rm IR}} & \dcolhead{M_{\rm dust}} & \dcolhead{\rm SFR_{\rm [C\,II]}} & \dcolhead{\rm SFR_{\rm dust}}\\
\colhead{} & \dcolhead{\rm[L_{\odot}]} & \dcolhead{\rm 10^7\, [M_{\odot}]} &  \dcolhead{\rm [M_{\odot}\,\rm yr^{-1}]} & \dcolhead{\rm [M_{\odot}\,\rm yr^{-1}]}}\tablecolumns{5}
\startdata
PSO\,J004+17 & 12.36\pm0.01 & 10.3\pm1.6 & 20 & 340 \\
PSO\,J011+09 & 12.56\pm0.01 & 16.5\pm2.5 & 30 & 540 \\
PSO\,J056--16 & 11.65\pm0.01 & 2.0\pm0.3 & 1 & 70 \\
PSO\,J158--14 & 12.98\pm0.01 & 43.2\pm6.6 & 230 & 1420 \\
SDSS\,J1143+3808 & 11.12\pm0.01 & 0.6\pm0.1 & - & 20 \\
PSO\,J239--07 & 11.81\pm0.01 & 2.9\pm0.5 & 20 & 100 \\
PSO\,J261+19 & <11.2 & <0.7 & - & <20 \\
PSO\,J265+41 & 12.99\pm0.01 & 44.7\pm6.9 & 1710 & 1470 \\
CFHQS\,J2100--1715 & 12.16\pm0.01 & 6.5\pm1.2 & 170^b & 210 \\
CFHQS\,J2229+1457 & 12.18\pm0.01 & 6.9\pm3.8 & 60^c & 230 \\
PSO\,J359--06 & 12.28\pm0.01 & 8.7\pm1.3 & 50 & 290 \\
 \enddata
\tablecomments{The columns show the name of the quasar, the estimated IR luminosity, and dust mass, as well as the inferred SFR of the quasars' host galaxies by means of the \cii\ emission line or the dust continuum. The $\rm SFR_{\rm [CII]}$ for CFHQS\,J2100--1715 and CFHQS\,J2229+1457 were derived by (\textit{a}) \citet{Decarli2018} and (\textit{b}) \citet{Willott2015}, respectively. }
\end{deluxetable*}

\subsection{Notes on Individual Objects}

In Fig.~\ref{fig:RpM} we show all proximity zone measurements that are not prematurely truncated or potentially contaminated by BAL features, as a function of the quasars' absolute magnitude $M_{1450}$. All quasars in our data sample show smaller proximity zone sizes than the expected average given their magnitude, which results from our selection criteria aiming to target young quasars. Five quasars that show no associated absorption systems or broad absorption lines, exhibit extremely small proximity zones with $R_{p,\,\rm corr}<2$~pMpc. 
These five quasars plus two from our previous study \citep{Eilers2017a}, which are marked with boxes in Fig.~\ref{fig:RpM}, indicate very short quasar lifetimes, i.e. $t_{\rm Q}\lesssim 10^5$~yr, which will be analyzed in more detail in \citetalias{E20prep}. 

Fig.~\ref{fig:LM} shows the bolometric luminosity as a function of the quasars' black hole masses. Our new measurements are compared to a low redshift quasar sample of $\gtrsim 75,000$ objects from the SDSS Data Release 7 (DR7) \citep{Shen2011, Wang2015}, as well as to several other $z\gtrsim5.8$ quasars \citep{Jiang2007, Willott2010, DeRosa2011, DeRosa2014, Wu2015, Mazzucchelli2017, Banados2018}. The quasars with very small proximity zones marked with red boxes do not populate a special region in the parameter space. 

\subsubsection{Quasars with Particularly Small Proximity Zones}\label{sec:newYQSOs}

\subsubsection*{PSO\,J004+17}
This object has a very small proximity zone of $R_p=1.16\pm0.15$~pMpc ($R_{p,\,\rm corr}=1.71\pm0.22$~pMpc). Although we find a high-ion absorption system within the quasar's proximity zone, it is not optically thick, and thus could not have prematurely truncated the proximity zone (see Fig.~\ref{fig:abs_J004}). The redshift of the absorption system coincides with the quasar's systemic redshift and might thus be due to the circumgalactic medium of the quasar's host galaxy. The dust continuum map shown in Fig.~\ref{fig:cont_map} reveals two other continuum sources in the vicinity of this quasar. 

\subsubsection*{VDES\,J0330--4025}
This object exhibits a small proximity zone of $R_p=1.68^{+0.62}_{-0.38}$~pMpc ($R_{p,\,\rm corr}=2.11^{+0.78}_{-0.48}$~pMpc) with no signs of close absorption systems (see Fig.~\ref{fig:abs_J0330}). It also has the largest black hole mass within our sample, i.e. $M_{\rm BH}=(4.96\pm0.51)\times 10^9\,M_{\odot}$. 

\subsubsection*{PSO\,J158--14}
This quasar, discovered both by \citet{Chehade2018} and Ba\~nados et al. in prep., shows a small proximity zone of $R_p=1.91\pm0.14$~pMpc ($R_{p,\,\rm corr}=1.63\pm0.12$~pMpc). Interestingly, the quasar has a very strong dust continuum emission, i.e. $F_{\rm cont}=3.46\pm0.02\,\rm mJy$. The derived star formation rate of approximately $1420\,M_{\odot}\,\rm yr^{-1}$ suggests the presence of a coeval starburst with the SMBH growth. Furthermore, we estimate a large bolometric luminosity of $\log L_{\rm bol}/\rm erg\,s^{-1} = 47.31$ and a high Eddington ratio of $\lambda_{\rm Edd}=1.01\pm0.08$.
The \mgii\ emission line is highly blueshifted with respect to the systemic redshift of the quasar, i.e. $\Delta v=-673\pm49\,\rm km\,s^{-1}$, suggesting strong internal motions within the BLR. 

\subsubsection*{CFHQS\,J2100--1715}
This quasar exhibits the smallest proximity zone $R_p=0.37\pm0.14$~pMpc ($R_{p,\,\rm corr}=0.66\pm0.25$~pMpc) detected to date with no signs of contamination from associated absorption systems (see Fig.~\ref{fig:absorption2}). \citet{Decarli2017} reported the detection of a companion galaxy at a projected separation of $\sim 60$~kpc, suggesting that the two objects might be at an early stage of interaction. Its spectrum shows a very red spectral slope %with $f_\lambda\propto\lambda^{-0.17}$
\citep{Willott2009}. 

\subsubsection*{CFHQS\,J2229+1457}
This quasar's small proximity zone has been confirmed in this study, i.e. $R_p=0.47\pm0.14$~pMpc ($R_{p,\,\rm corr}=1.12\pm0.33$~pMpc). Its X-Shooter spectrum does not reveal any associated absorption systems that could truncate or contaminate its proximity zone, although the SNR of the data redwards of the \lya\ emission line is still very low (see Fig~\ref{fig:abs_J2229}). Note that the black hole mass measurement for this object should be taken with caution, because the \mgii\ emission line falls on top of a telluric feature. Our measurement differs from a previous measurement by \citet{Willott2010} who made use of a lower resolution spectrum ($R\approx 520$) observed with NIRI/Gemini by more than one order of magnitude. 

\subsubsection{Remaining Objects}

\subsubsection*{PSO\,J011+09}
This object the second highest-redshift quasar in our data sample. Its \mgii\ emission line is highly blueshifted, i.e. $\Delta v=-1021\pm143\,\rm km\,s^{-1}$, with respect to the \cii\ systemic redshift. 

\subsubsection*{PSO\,J056--16}
As shown in Fig.~\ref{fig:absorption} this quasar's proximity zone has been prematurely truncated due to a pDLA along our line-of-sight. % \citep[][]{Davies2019c}. 
Thus, we exclude this object from any further analysis of its proximity zone. The estimated large Eddington ratio of $\lambda_{\rm Edd}=1.26\pm0.08$ indicates that this quasar has a high accretion rate.

\subsubsection*{VDES\,J0323--4701}
It has been speculated that this quasar as well as VDES\,J0330--4025 might lie in an overdense region of the universe, since they are located within $10$ degrees on the sky \citep{Reed2017}. The new, very similar redshift estimates for both objects based on their \mgii\ emission lines of $z_{\rm Mg\,II}\approx 6.241$ with a very small velocity difference of $\Delta v\approx 50\rm\,km\,s^{-1}$ supports this hypothesis. The estimated Eddington ratio is very high, i.e. $\lambda_{\rm Edd}=1.76\pm1.24$. However, the NIR spectrum of this object and in particular the \mgii\ emission line have a very low SNR and thus these estimates have large uncertainties and should be taken with caution. 

\subsubsection*{SDSS\,J1143+3803}
Based on the new redshift estimate from the CO(6--5) emission line ($z_{\rm CO(6-5)}\approx5.8366$), which is significantly higher than the preliminary estimate from the \lya\ emission line ($z_{\rm Ly\alpha}\approx5.805$), this quasar has the largest proximity zone ($R_p=3.93\pm0.63$~pMpc) in our sample. Unfortunately, the end of its zone falls right in between the two DEIMOS detectors, which are separated by a $\Delta \lambda =11$~{\AA} wide gap. Thus, we adopted the middle of the detector gap as the best estimate of the proximity zone size and the width of the gap as the uncertainty on $R_p$. 

\subsubsection*{PSO\,J239--07}
Although this quasar exhibits a very small proximity zone, i.e. $R_p=1.29\pm0.14$~pMpc, the broad absorption line features detected in its optical/NIR spectrum might have contaminated its zone. Thus, we exclude this object from any further analysis. The dust continuum map shown in Fig.~\ref{fig:cont_map} shows two other continuum sources in the vicinity of this quasar. 

\subsubsection*{PSO\,J261+19}
This quasar is the highest redshift object in our sample. We could not detect any line nor continuum emission associated with its host galaxy (see Fig.~\ref{fig:postage} and Fig.~\ref{fig:cont_map}), indicating that its emission is very faint, i.e. $F_{\rm cont}<0.05\rm\,mJy$. %Its NIR spectrum suggests a high Eddington ratio of $\lambda_{\rm Edd}=1.07\pm0.38$.

\subsubsection*{PSO\,J265+41}
This object has been discovered by Ba\~nados et al. in prep. It is a BAL quasar, and has thus been eliminated from our analysis about proximity zones, since its absorption features might contaminate the proximity zone. This quasar shows a very bright \cii\ line ($F_{\rm line} = 9.20\pm0.50\,\rm Jy\,km\,s^{-1}$), as well as a bright dust continuum emission ($F_{\rm cont} = 3.61\pm0.07$), suggesting a highly star-forming (${\rm SFR}\gtrsim 1470\,\rm M_{\odot}\,yr^{-1}$) quasar host galaxy.

\subsubsection*{PSO\,J359--06}
This object's NIR spectrum suggests a high Eddington ratio of $\lambda_{\rm Edd}=0.90\pm0.05$. 

\subsection{Fraction of Young Quasars}

In our previous study we discovered three young quasars in a parent sample of $31$. This implies a fraction of young quasars within the quasar population at large of $f_{\rm young}\approx3/31\approx 10\%$ \citep{Eilers2017a}. For a majority of objects in this sample we had precise redshift estimates and good spectroscopic data, and hence the young population within this data set is likely to be complete. 

Assuming that the five objects mentioned in \S~\ref{sec:newYQSOs} all have short quasar lifetimes we now know a total of seven young objects. Please be reminded that one of the here analyzed objects was already previously identified to have a short quasar lifetime \citep{Eilers2017a}. Given a total quasar sample of $153$ objects, i.e. the combined $122$ quasars from the parent sample of this study (\S~\ref{sec:quasar_sample}) and the $31$ objects from \citet{Eilers2017a}, we obtain an estimated fraction of young quasars $f_{\rm young}\approx 7/153 \approx 5\%$. However, this estimate likely represents a conservative lower limit, since there might still be more quasars with short lifetimes within the remaining sample for which we did not conduct follow-up observations and thus no precise redshift estimates exist to date. Additionally, quasars exhibiting BAL features, which can make up to $\sim 40\%$ of the quasar population \citep[e.g.][]{Allen2011}, could also be young, but we will not be able to estimate their lifetime by means of their proximity zone sizes. 

Thus, we conclude that the fraction of young quasars within the whole quasar population in the early universe is $5\%\lesssim f_{\rm young}\lesssim 10\%$. 

\section{Summary \& Conclusions}\label{sec:summary}

We perform a multi-wavelength analysis to systematically detect and characterize high-redshift quasars that are likely to be very young, as indicated by their small proximity zones. We analyze $13$ quasars at $5.8 \lesssim z \lesssim 6.5$, and determine precise redshift estimates by means of their \cii\ or $\rm CO\,(6-5)$ emission lines arising from the cold gas of the host galaxy observed with ALMA and NOEMA, or based on their \mgii\ emission line from the NIR spectra observed with VLT/X-Shooter, if no sub-mm data is available. These new redshift estimates allow us to precisely measure the size of the proximity zones of the quasars, which we will use in a subsequent follow-up paper to determine their lifetimes \citepalias{E20prep}. Additionally, the deep optical and NIR spectra we obtained from VLT/X-Shooter and Keck/DEIMOS for the quasar sample allow us to exclude a contamination of the proximity zone due to associated absorption systems or broad absorption lines, and enable measurements of the black hole masses, as well as the Eddington ratio of their accretion. 

The main results of this study are:
\begin{enumerate}
    \item We find five quasars (PSO\,J004+17, VDES\,J0330--4025, PSO\,J158--14, CFHQS\,J2100--1715, and CFHQS\,J2229+1457) that exhibit particularly small proximity zones, i.e. $R_{p,\,\rm corr}\lesssim 2$~pMpc, and thus likely indicate very short quasar lifetimes, i.e. $t_{\rm Q}\lesssim 10^5$~yr. One of these five objects, CFHQS\,J2229+1457, has previously been identified as a very young quasar \citep{Eilers2017a}. 
    \item The quasar CFHQS\,J2100--1715 exhibits the smallest proximity zone detected to date, i.e. $R_{p,\,\rm corr}=0.66\pm0.25$~pMpc. Additionally, the detection of a companion galaxy \citep{Decarli2017} might indicate that this system is at an early stage of interaction. 
    \item Our previous work revealed three young quasars in a sample of $31$ objects \citep{Eilers2017a}. For this study, we analyzed $122$ quasar spectra, and chose the most promising $13$ candidates to follow-up. We discover five young quasars in this data set, one of which was previously known. This allows us to constrain the fraction of young quasars within the high-redshift quasar population at large to be $5\%\lesssim f_{\rm young}\lesssim 10\%$.
    \item We determine the spectral properties of the quasars in our sample, such as black hole masses, the velocity shifts of the emission lines, and the Eddington ratios of the accretion of the SMBHs. For three objects in the sample we measure large Eddington ratios, i.e. $\lambda_{\rm Edd}\gtrsim 1$, which indicate high mass accretion rates at the Eddington limit. The estimated black hole masses, derived from fitting the single-epoch \mgii\ region, vary between $M_{\rm BH}\approx 3\times10^8 - 5\times 10^9\,M_{\odot}$. 
    \item We measure dust continuum fluxes and find two particularly bright quasars, i.e. PSO\,J158--14 (which is likely to have a very short lifetime) and PSO\,J261+41, with $F_{\rm cont}=3.46\pm 0.02\,\rm mJy$ and $F_{\rm cont}=3.61\pm 0.07\,\rm mJy$, respectively. The inferred high star formation rates of $\gtrsim 1400\,M_{\odot}\,\rm yr^{-1}$ suggest the presence of a starburst within their host galaxies coeval to the SMBH growth. These quasars represents ideal targets for high-resolution imaging with ALMA to study star formation and the ISM in the very early universe, and search for any signs of mergers, outflows, and feedback. 
\end{enumerate}

Our analysis presents a first step towards a systematic study of the lifetimes of high-redshift quasars based on their proximity zone sizes. In future work we will determine the lifetime estimates for this quasar sample, as well as a statistical estimate of the lifetime of the quasar population at large by means of the estimated fraction of young quasars. This will enable us to analyze the evolution of quasar and host galaxy properties with the quasar lifetime, and further study the accretion behaviour of SMBHs in the early universe. 

\software{CASA \citep{McMullin2007}, GILDAS (\url{http://www.iram.fr/IRAMFR/GILDAS}), PypeIt (DOI: 10.5281/zenodo.3506873), numpy \citep{numpy}, scipy \citep{scipy}, matplotlib \citep{matplotlib}, astropy \citep{astropy}}

\acknowledgements

The authors would like to thank the anonymous referee for very thorough and constructive feedback. Furthermore, we would like to thank Sarah Bosman for helpful comments on the manuscript, as well as Michael Rauch for sharing his data. 

This paper makes use of the following ALMA data: ADS/JAO.ALMA\#2017.1.00332.S. ALMA is a partnership of ESO (representing its member states), NSF (USA) and NINS (Japan), together with NRC (Canada), MOST and ASIAA (Taiwan), and KASI (Republic of Korea), in cooperation with the Republic of Chile. The Joint ALMA Observatory is operated by ESO, AUI/NRAO and NAOJ.

This work is based on observations carried out under project number W17EQ and W18EF with the IRAM NOEMA Interferometer [30m telescope]. IRAM is supported by INSU/CNRS (France), MPG (Germany) and IGN (Spain). RD and ACE thank Charlene Lefevre for her support in the IRAM data calibration. 

This work is based on observations collected at the European Organisation for Astronomical Research in the Southern Hemisphere under ESO programmes 096.A-0418, 097.B-1070, 098.B-0537, and 101.B-02720. 

Some of the data presented in this paper were obtained at the W.M. Keck Observatory, which is operated as a scientific partnership among the California Institute of Technology, the University of California and the National Aeronautics and Space Administration. The Observatory was made possible by the generous financial support of the W.M. Keck Foundation. 

The authors wish to recognize and acknowledge the very significant cultural role and reverence that the summit of Mauna Kea has always had within the indigenous Hawaiian community. We are most fortunate to have the opportunity to conduct observations from this mountain. 

ACE and F. Wang acknowledge support by NASA through the NASA Hubble Fellowship grant $\#$HF2-51434 and $\#$HST-HF2-51448.001-A awarded  by  the  Space  Telescope  Science  Institute,  which  is  operated  by  the   Association  of  Universities for  Research  in  Astronomy,  Inc.,  for  NASA,  under  contract  NAS5-26555. 

BPV and MN acknowledge funding through the ERC grant ``Cosmic Gas''. 

This publication has received funding from the European Union’s Horizon 2020 research and innovation programme under grant agreement No 730562 [RadioNet].

\appendix

\section{Dust Continuum Maps}\label{app:continuum}

\begin{figure*}[!h]
\centering
\includegraphics[width=.95\textwidth]{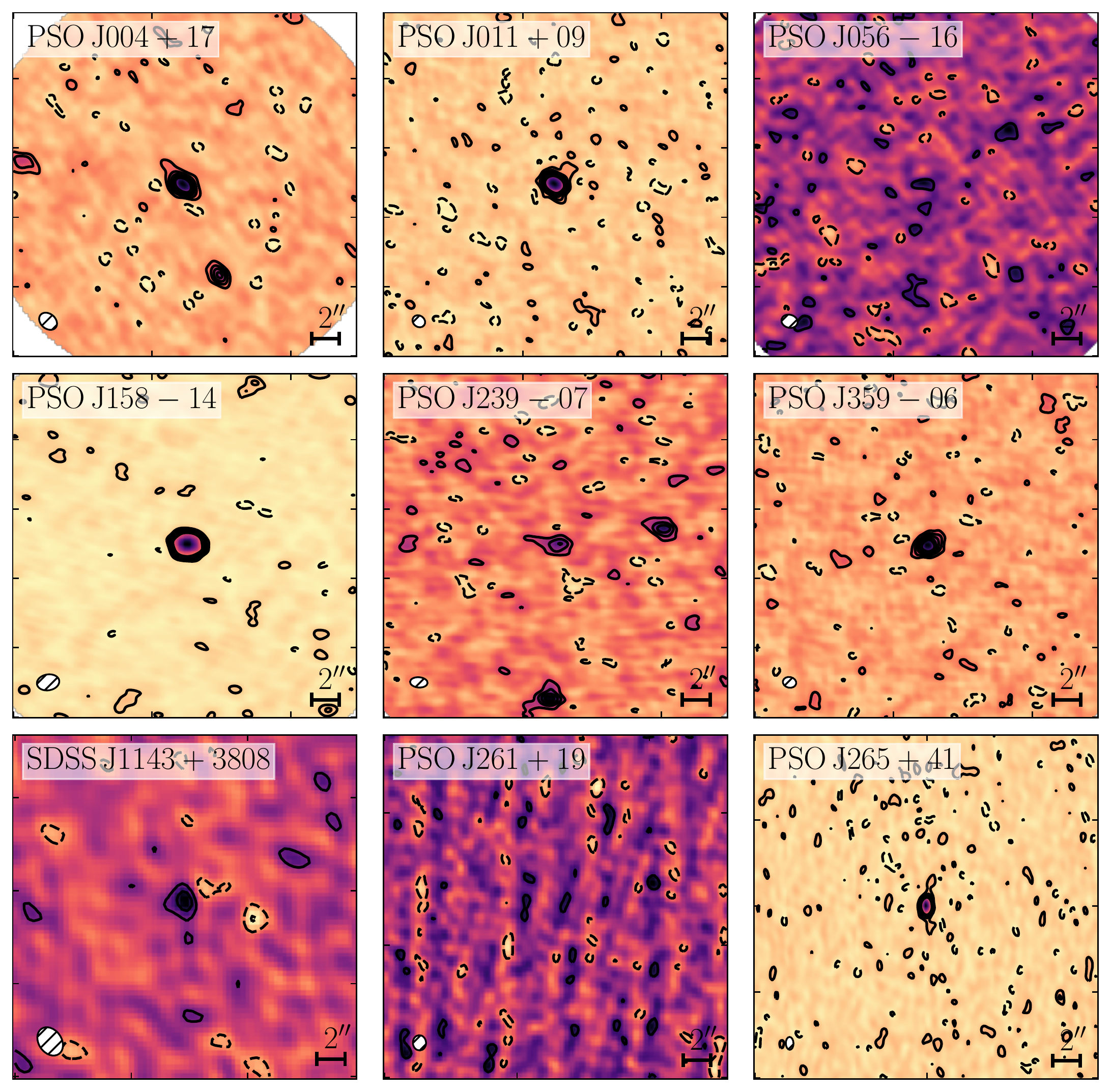}
\caption{\textbf{Dust continuum maps.} The six quasars in the top and middle row show ALMA observations, whereas the bottom row shows our NOEMA data. Each panel is $25\arcsec\times 25\arcsec$ in size. Solid and dashed lines show the $\pm 2, 4, 6, 8, 10\sigma$ isophotes. \label{fig:cont_map}} 
\end{figure*}

The dust continuum maps constructed from the two ALMA bandpasses without the \cii\ emission line, or the line-free channels from NOEMA are shown in Fig.~\ref{fig:cont_map}. We find two continuum sources in close (projected) vicinity to the quasars PSO\,J004+17 and PSO\,J239--07. 

\section{\ion{Mg}{2} emission line fits}\label{app:mgii}
\begin{figure*}[!h]
\centering
\includegraphics[width=.9\textwidth]{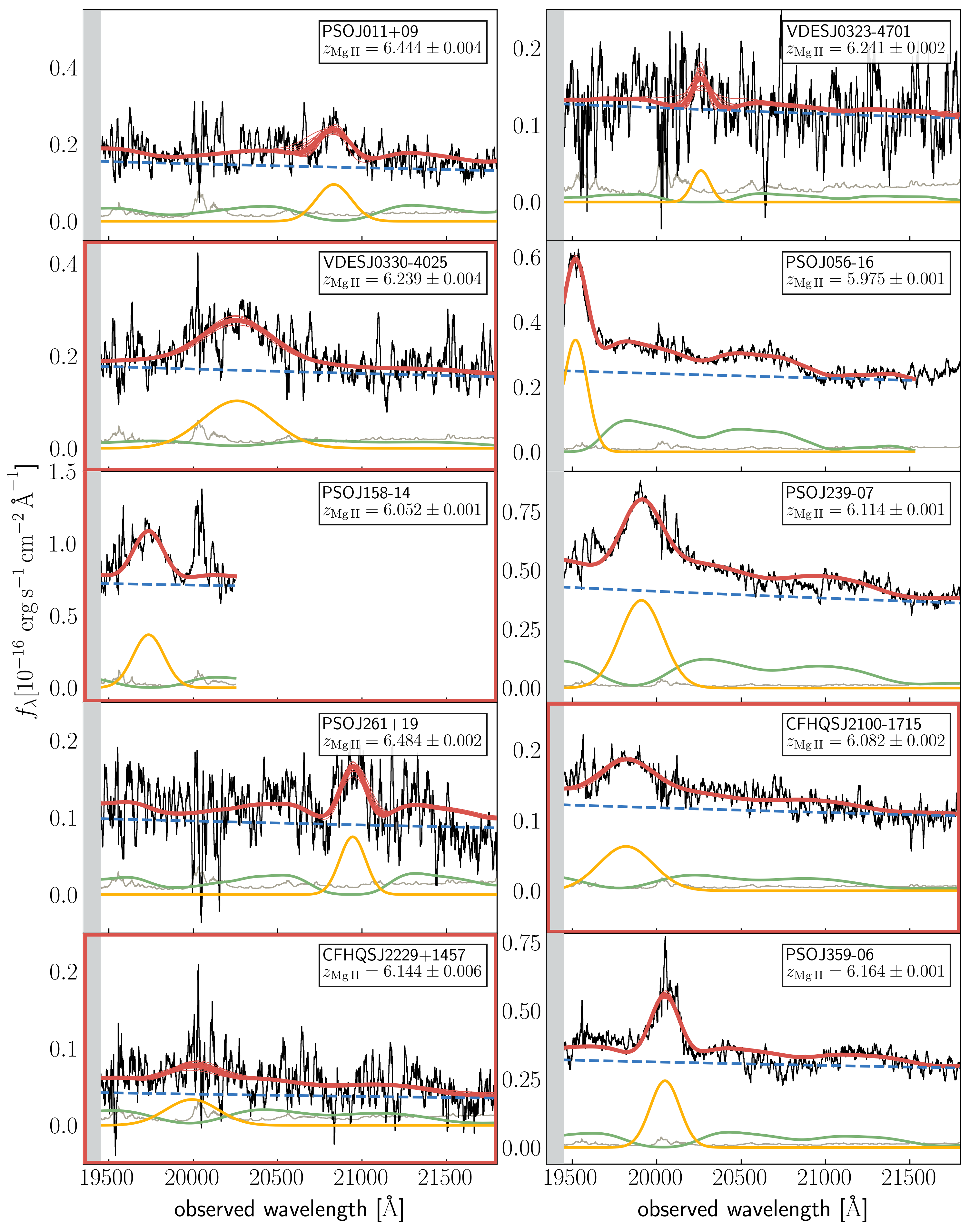}
\caption{\textbf{Best fits to the \mgii\ emission lines.} We show all quasars in our sample that have NIR spectral coverage and their \mgii\ line does not fall into a region of high telluric absorption. The spectra around \mgii\ at $\lambda_{\rm rest}=2798$~{\AA} and the corresponding noise vector have been inverse-variance smoothed with a $20$ pixel filter and are shown in black and grey, respectively. The fit to the quasar emission (red curve), as well as its individual components are shown as the colored curves, i.e. a power-law continuum (blue dashed), the smoothed iron template (green), and a Gaussian for the \mgii\ emission line. Regions with large telluric absorption have been masked by the grey regions. The faint red lines show draws from the posterior distribution. The red frames show quasars that exhibit very small proximity zones and short lifetimes. \label{fig:mgii}} 
\end{figure*}

In order to measure the black hole masses of the quasar, we fit the width of the \mgii\ emission line as described in \S~\ref{sec:BH}. The best fits to the \mgii\ emission lines are shown in Fig.~\ref{fig:mgii}. 

\section{Quasar Continuum Normalization}\label{app:cont}

\begin{figure*}[!t]
\centering
\includegraphics[width=.95\textwidth]{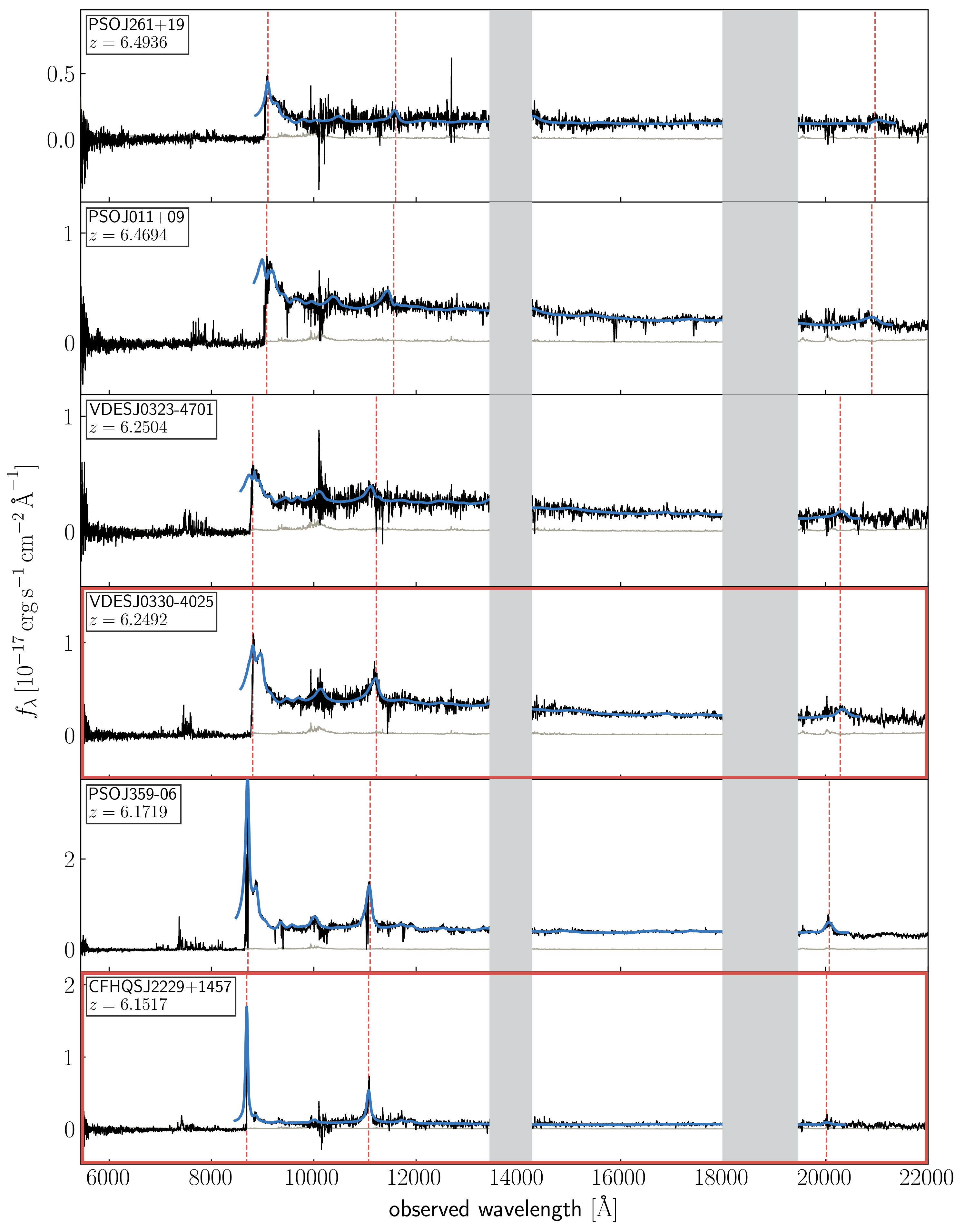}
\caption{Same as Fig.~\ref{fig:spectra_a} with quasar continuum estimates, shown in blue. \label{fig:spectra_cont_a}} 
\end{figure*}

\begin{figure*}[!t]
\centering
\includegraphics[width=.95\textwidth]{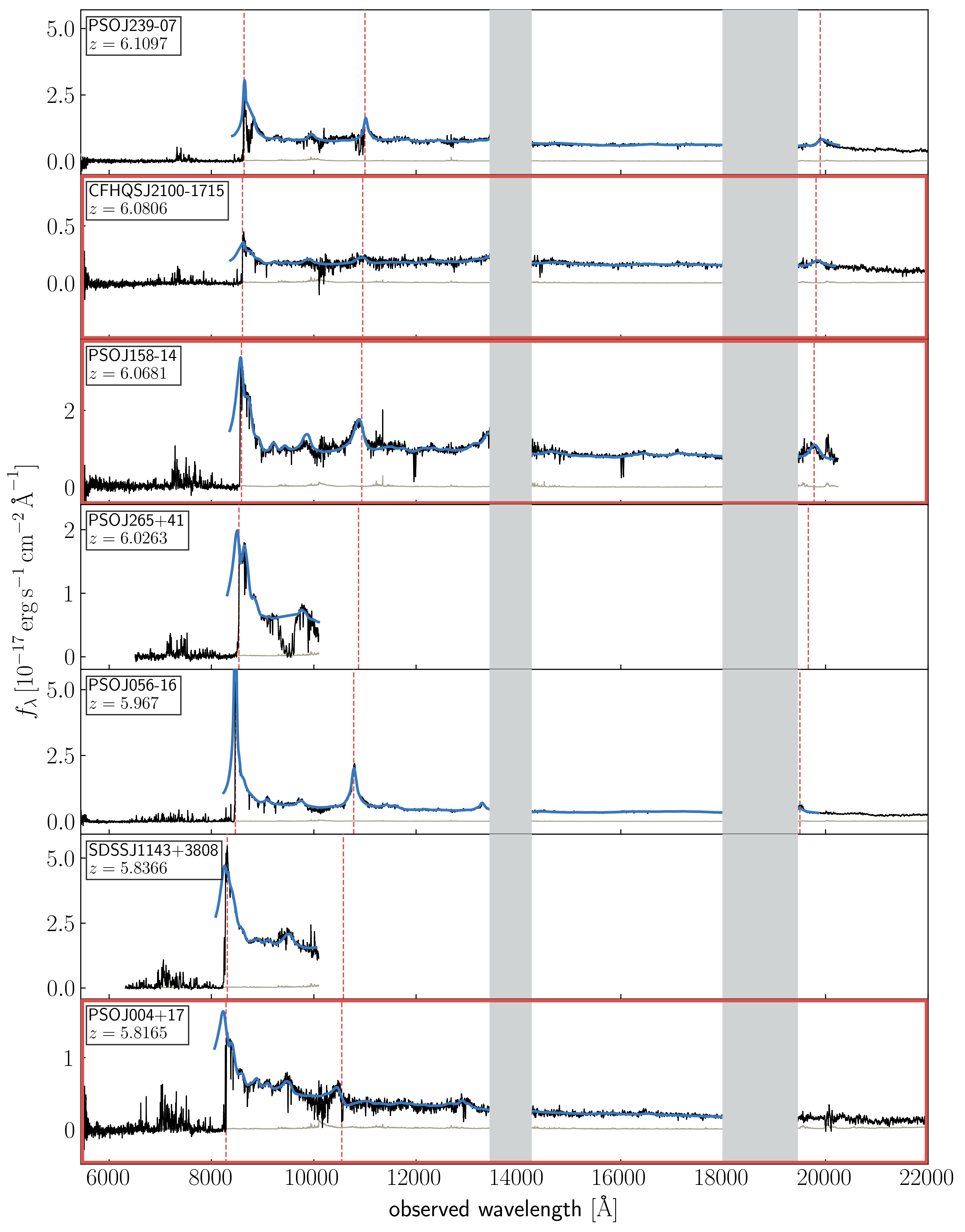}
\caption{Same as Fig.~\ref{fig:spectra_b} with quasar continuum estimates, shown in blue. \label{fig:spectra_cont_b}} 
\end{figure*}

In \S~\ref{sec:cont} we described our method to estimate the intrinsic quasar continua. We show the best estimate for each quasar spectrum in Fig.~\ref{fig:spectra_cont_a} and Fig.~\ref{fig:spectra_cont_b}. 

\subsection{Influence of Continuum Uncertainties on Proximity Zone Measurements}\label{app:continuum_uncertainties}

There are uncertainties on the estimated quasar continua bluewards of the \lya\ emission line due to uncertainties intrinsic to the PCA method we apply, i.e. stochastic uncertainties in the relationship between red-side and blue-side features, as well as the inability of the PCA model to exactly reproduce a given spectrum. Following \citet{Davies2018}, we estimate that the mean of the error (i.e. the bias) is $\epsilon_C\approx1\%$, whereas the uncertainty of the prediction can be up to $\sigma_{\epsilon_C}\approx 10\%$. 

\begin{figure*}[!h]
\centering
\includegraphics[width=\textwidth]{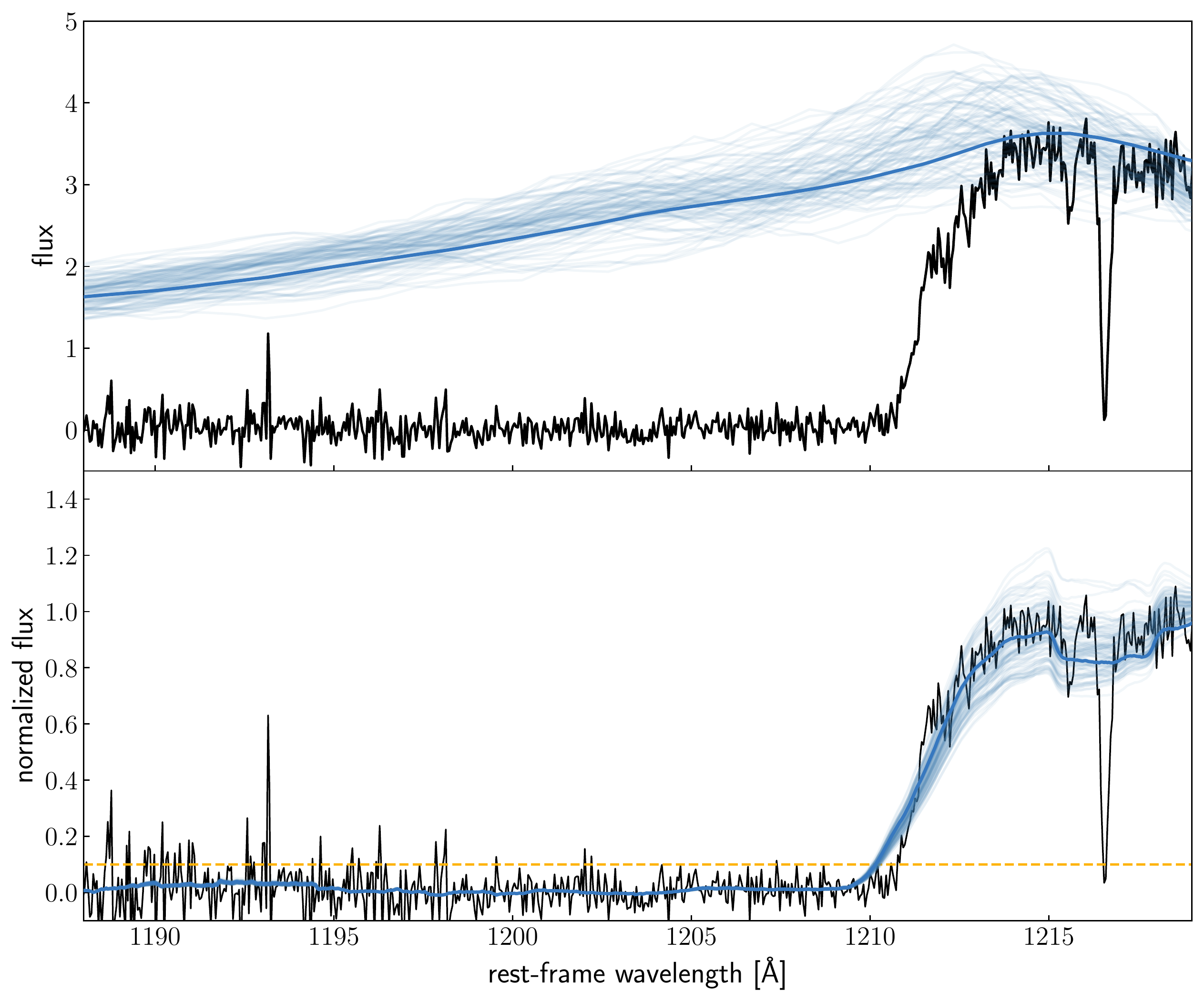}
\caption{Example for analyzing the effects of the quasar continuum uncertainties on the proximity zone measurements. \textit{Top:} Spectrum of PSO\,158--14 (black) with $100$ draws from the quasar continuum estimate including Gaussian noise (blue). \textit{Bottom:} Continuum normalized quasar spectrum and the smoothed flux from the different draws of the continuum estimate, showing that the influence of continuum uncertainties is very small on the location of $R_p$, where the smoothed flux drops below the $10\%$-level (yellow). \label{fig:continuum_uncertainties}} 
\end{figure*}

In order to test the influence of these uncertainties in the quasar continuum estimate on the measurements of the proximity zones we draw samples of the continuum with Gaussian uncertainties added according to $\sigma_{\epsilon_C}$, as shown in Fig.~\ref{fig:continuum_uncertainties} for an example spectrum. We then calculate the proximity zones for each draw as described in \S~\ref{sec:rp}, and estimate the error on $R_p$ due to uncertainties in the continuum estimate, i.e.\ $\Delta R_p\approx 0.01-0.03$~pMpc, for all quasars for which NIR spectral coverage was available to construct the PCA model. For quasars without NIR coverage we had to use a truncated PCA model, which results in slightly larger uncertainties and thus the error on $R_p$ increases to $\Delta R_p \approx 0.05-0.25$~pMpc. These uncertainties are still small compared to the $R_p$ measurements. 

\section{Absorption Systems}\label{app:abs}

\begin{figure*}[!h]
\centering
\includegraphics[width=\textwidth]{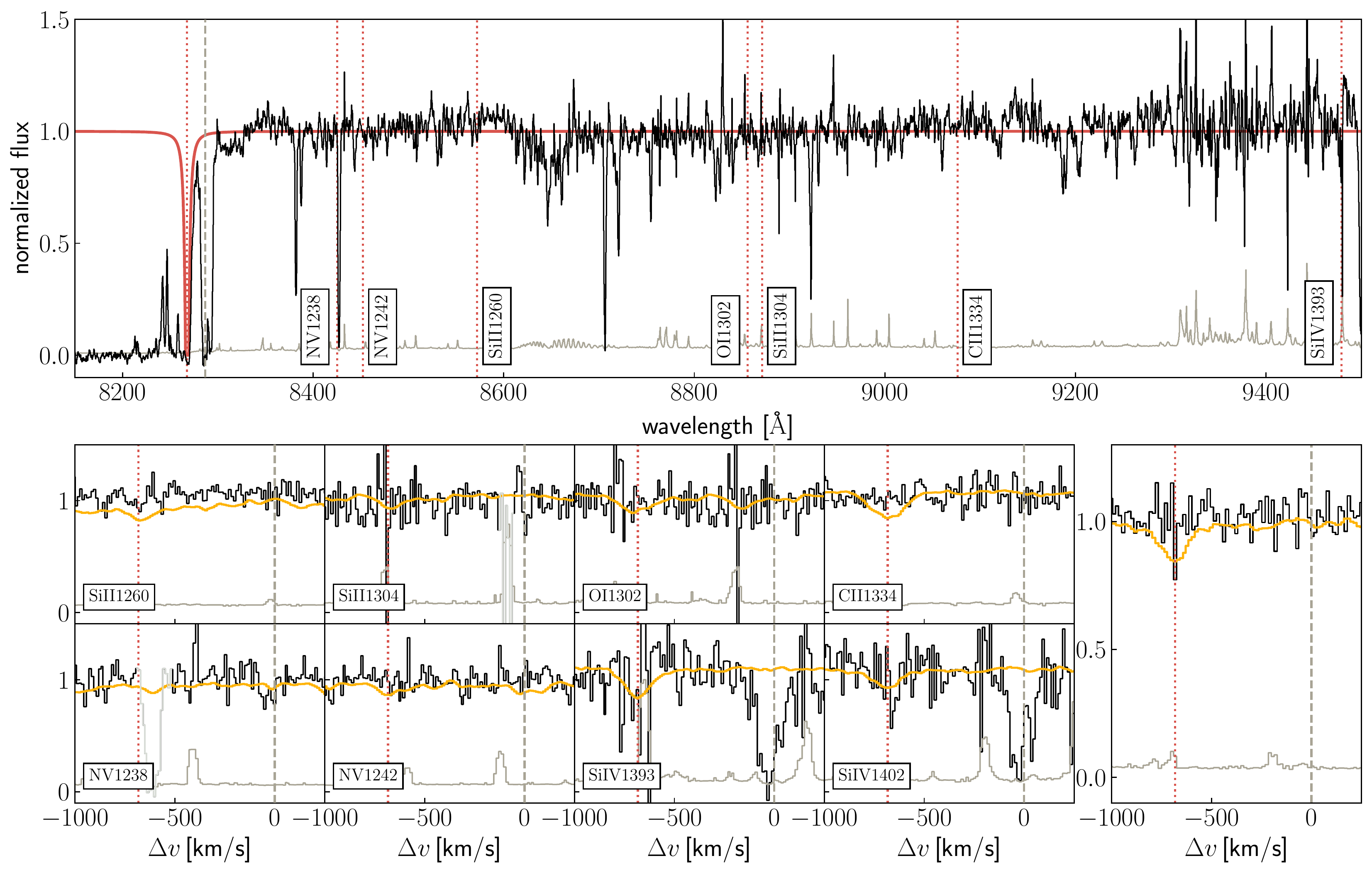}
\caption{Same as Fig.~\ref{fig:absorption2} but for PSO\,J004+17. For this object we also show the locations of the high-ionization absorption, since these show evidence for an absorption system at the systemic redshift of the quasar (grey dashed lines). Wavelength regions (shown in light grey) around very prominent skylines, as well as around low-redshift absorption systems have been masked to avoid biases in the stack. \label{fig:abs_J004}} 
\end{figure*}

\begin{figure*}[!h]
\centering
\includegraphics[width=\textwidth]{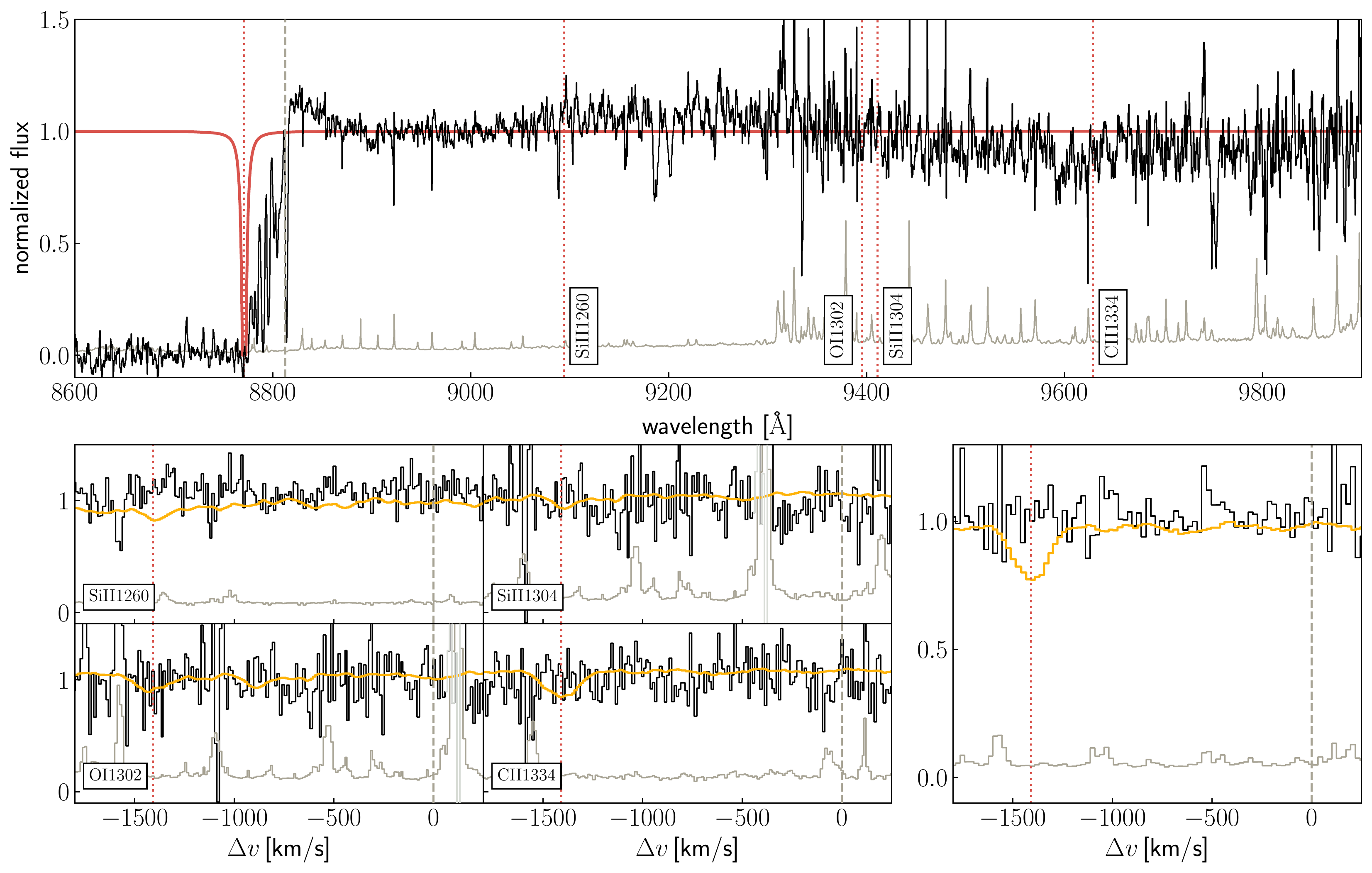}
\caption{Same as Fig.~\ref{fig:absorption2} but for VDES\,J0330--4025. Wavelength regions around very prominent skylines (shown in light grey) have been masked to avoid biases in the stack. \label{fig:abs_J0330}} 
\end{figure*}

\begin{figure*}[!h]
\centering
\includegraphics[width=\textwidth]{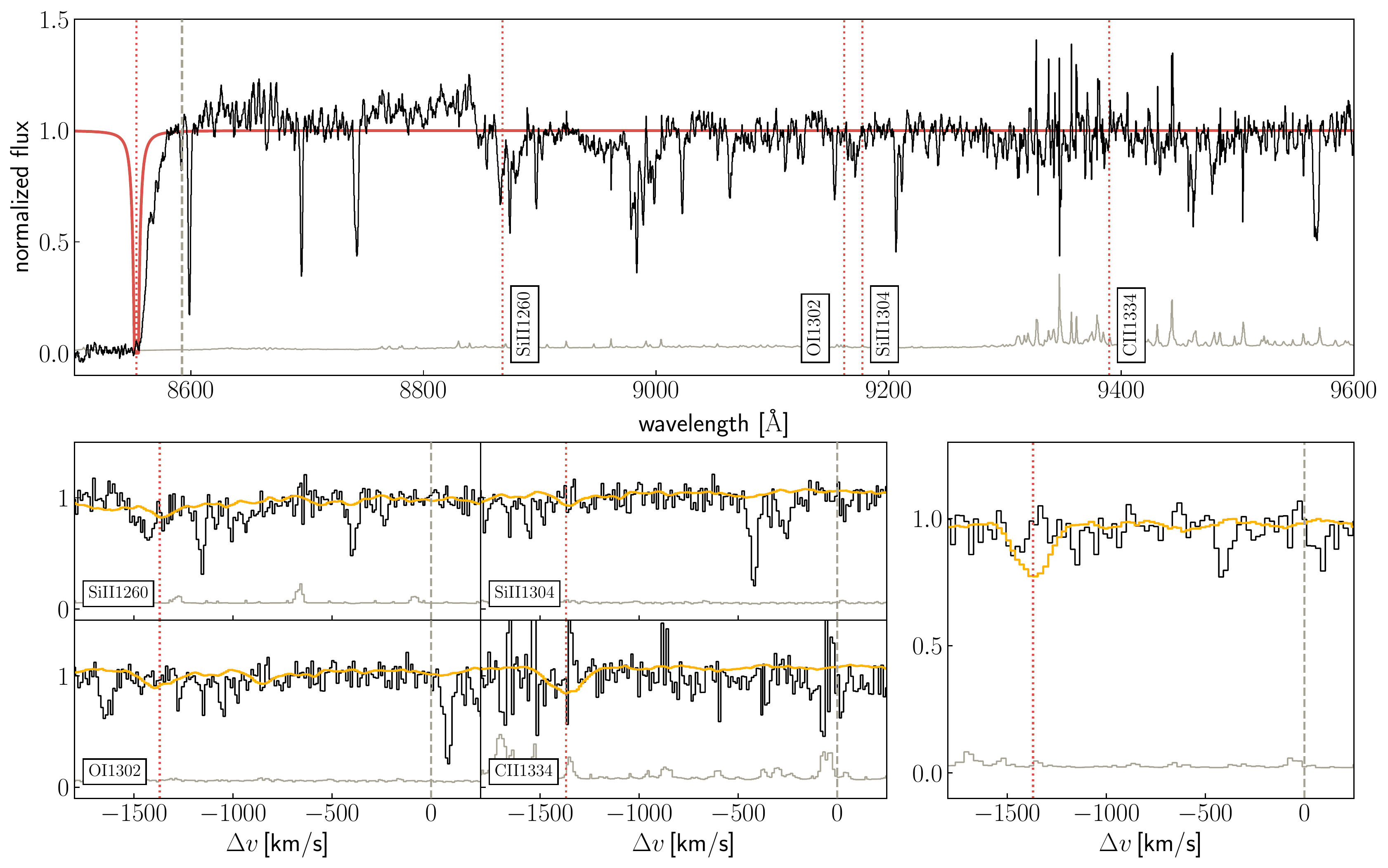}
\caption{Same as Fig.~\ref{fig:absorption2} but for PSO\,J158--14. \label{fig:abs_J158}} 
\end{figure*}

\begin{figure*}[!h]
\centering
\includegraphics[width=\textwidth]{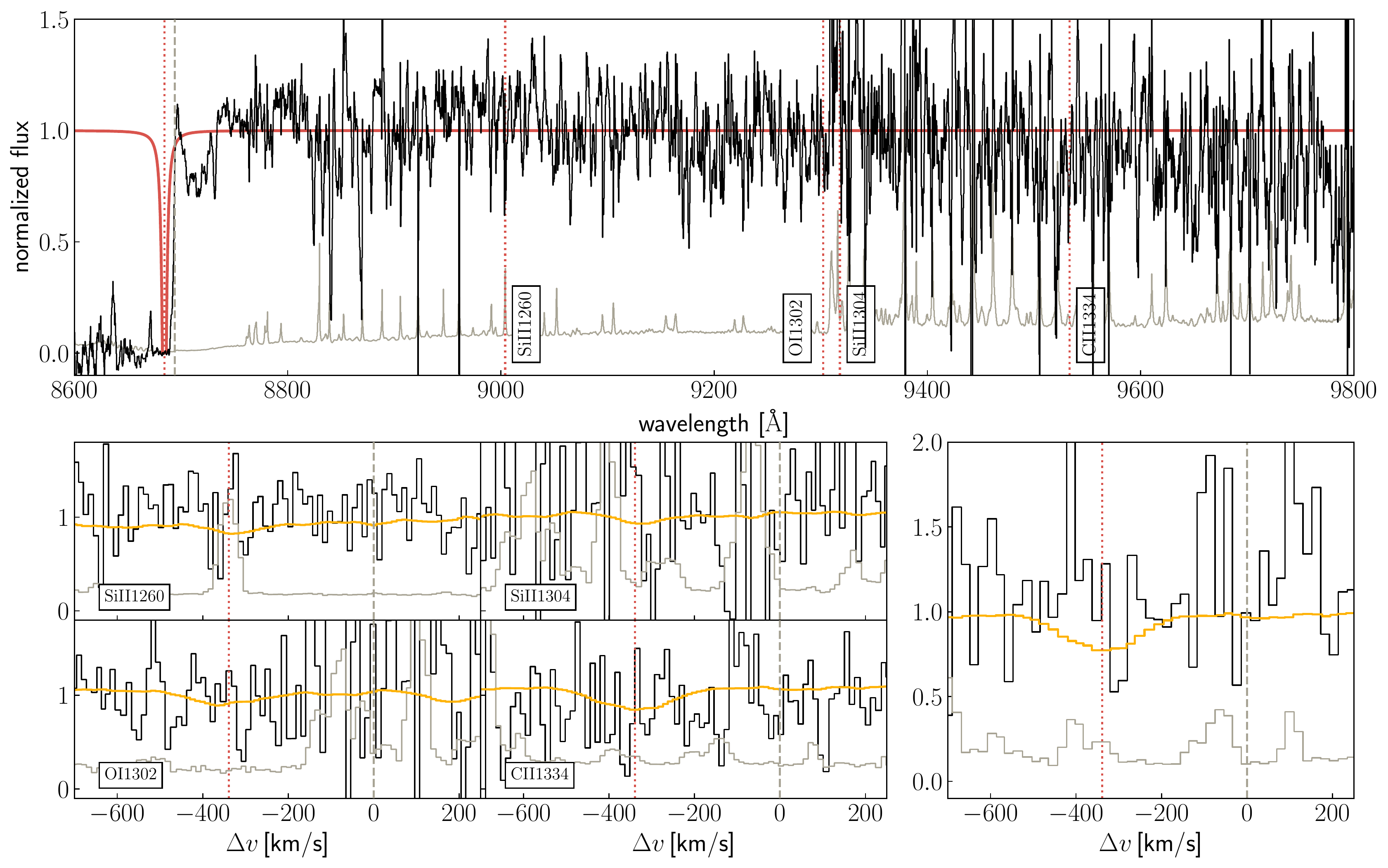}
\caption{Same as Fig.~\ref{fig:absorption2} but for CFHQS\,J2229+1457. \label{fig:abs_J2229}} 
\end{figure*}

In Fig.~\ref{fig:abs_J004} to \ref{fig:abs_J2229} we show the spectra of all quasars with very small proximity zones and a hypothetical absorption system. We do not see any evidence for a premature truncation of an associated absorption system. 

However, the spectrum of PSO\,J004+17 shown in Fig.~\ref{fig:abs_J004} shows an associated absorption system located within the quasar's proximity zone directly at the systemic redshift of the quasar, i.e. $z_{\rm abs}\approx5.8165$, possibly due to the circumgalactic medium of the host galaxy itself. However, this system only shows high-ionization absorption lines, such as the doublets \ion{N}{5} at the rest-frame wavelengths $\lambda=1238$~{\AA} and $\lambda=1242$~{\AA}, 
and \ion{Si}{4} at $\lambda=1393$~{\AA} and $\lambda=1402$~{\AA}.
We do not find any evidence for low-ionization lines, indicating that the absorption system is unlikely to be self-shielding. Additionally, the spectrum shows clear flux transmission bluewards of the absorption system, which indicates that the proximity zone extends beyond this absorber. Thus we exclude a premature truncation of the quasar's proximity zone due to this absorption system. 

The spectrum of PSO\,J158--14 shows a potential absorption line close to the location of \ion{Si}{2} $\lambda\,1260$ of the hypothetical absorption system. However, we do not find evidence for any other low-ionization lines that could be associated with an absorption system, which should be present if there was indeed such a system, and thus conclude that the line likely belongs to a foreground absorber at lower redshift.

We note that the SNR of the spectrum of CFHQS\,J2229+1457 redwards of the \lya\ emission line is still very low due to the quasar's faint continuum emission. We do not see any evidence for an absorption system that might truncate the proximity zone (see Fig.~\ref{fig:abs_J2229}), however, the low data quality does not allow us to securely rule out this option.

\bibliography{literatur_hz}

\end{document}